\documentclass[fleqn,usenatbib]{mnras_tex_edited}
\usepackage{newtxtext,newtxmath}

\usepackage[T1]{fontenc}
\usepackage{lscape}

\DeclareRobustCommand{\VAN}[3]{#2}
\let\VANthebibliography\thebibliography
\def\thebibliography{\DeclareRobustCommand{\VAN}[3]{##3}\VANthebibliography}

\usepackage{graphicx,epsf, epsfig}
\usepackage{bm}
\usepackage{longtable,tabularx}
\usepackage{color}
\definecolor{linkcolor}{rgb}{0.0,0.3,0.5}
\usepackage{array}
\usepackage{multirow}
\usepackage{rotating,array}
\usepackage[normalem]{ulem}
\usepackage[caption=false]{subfig} %
\usepackage{dcolumn}%
\usepackage{braket}
\usepackage{appendix}
\usepackage{comment}
\usepackage{siunitx}
\usepackage{orcidlink}
\usepackage{hyperref}

\definecolor{dodgerblue}{HTML}{1E90FF}

\newcommand{\bham}{School of Physics and Astronomy \&	Institute for Gravitational Wave Astronomy, University of Birmingham, Birmingham, B15 2TT, UK}
\newcommand{\milan}{Dipartimento di Fisica ``G. Occhialini'', Universit\'a degli Studi di Milano-Bicocca, Piazza della Scienza 3, 20126 Milano, Italy}
\newcommand{\infn}{INFN, Sezione di Milano-Bicocca, Piazza della Scienza 3, 20126 Milano, Italy}
\newcommand{\bjing}{Kavli Institute for Astronomy and Astrophysics, Peking University, Beijing 100871, China }

\newcommand{\logten}{\log_{10}}
\newcommand{\Msol}{\rm M_{\odot}}
\newcommand{\Mscale}{10^7 \Msol}
\newcommand{\Mc}{\mathcal{M}}
\newcommand{\Mstar}{\mathcal{M}_{\star}}
\newcommand{\ndot}{{\dot n}_0}
\newcommand{\alphaM}{\alpha_{\Mc}}
\newcommand{\betaz}{\beta_z}
\newcommand{\Ho}{H_0}
\newcommand{\OmegaL}{\Omega_{\Lambda}}
\newcommand{\OmegaM}{\Omega_{M}}
\newcommand{\Ayr}{A_{\rm yr}}

\def\araa{\rm{ARA\&A}}
\def\apj{\rm{Astrophys.~J.}}
\def\apjl{\rm{Astrophys.~J.~ Lett.}}
\def\apjs{\rm{ApJS}}
\def\aap{\rm{Astron.~Astrophys.}}

\def\mnras{\rm{Mon.~Not.~R.~Astron.~Soc.}}
\def\prd{\rm{Phys.~Rev.~D}}
\def\prl{\rm{Phys.~Rev.~Lett.}}

\title[LISA, PTAs, and massive black-hole binaries]
{Implications of pulsar timing array observations for LISA detections of massive  black hole binaries}

\author[N. Steinle {\it et al.}]{Nathan Steinle$\,$\orcidlink{0000-0003-0658-402X}$^{1}$\medskip\thanks{Contact e-mail: \href{mailto:nsteinle@star.sr.bham.ac.uk}{nsteinle@star.sr.bham.ac.uk}}, 
Hannah Middleton$\,$\orcidlink{0000-0001-5532-3622}$^{1}$, 
Christopher J.\ Moore$\,$\orcidlink{0000-0002-2527-0213}$^{1}$, 
Siyuan Chen$\,$\orcidlink{0000-0002-3118-5963}$^{2}$,\newauthor
Antoine Klein$\,$\orcidlink{0000-0001-5438-9152}$^{1}$,
Geraint Pratten$\,$\orcidlink{0000-0003-4984-0775}$^{1}$,
Riccardo Buscicchio$\,$\orcidlink{0000-0002-7387-6754}$^{3,4}$,
Eliot Finch$\,$\orcidlink{0000-0002-1993-4263}$^{1}$,\newauthor
Alberto Vecchio$\,$\orcidlink{0000-0002-6254-1617}$^{1}$
\\
$^{1}$\bham\\
$^{2}$ \bjing\\
$^{3}$ \milan\\
$^{4}$ \infn\\
}

\date{Accepted XXX. Received YYY; in original form ZZZ}

\pubyear{\the\year{}}

\begin{document}
\label{firstpage}
\pagerange{\pageref{firstpage}--\pageref{lastpage}}
\maketitle

\begin{abstract}
Pulsar timing arrays (PTAs) and the Laser Interferometer Space Antenna (LISA) will open complementary observational windows on massive black-hole binaries (MBHBs), i.e., with masses in the range $\sim 10^6 - 10^{10}\,\Msol$. 
While PTAs may detect a stochastic gravitational-wave background from a population of MBHBs, during operation LISA will detect individual merging MBHBs. 
To demonstrate the profound interplay between LISA and PTAs, we estimate the number of MBHB mergers that one can expect to observe with LISA by extrapolating direct observational constraints on the MBHB merger rate inferred from PTA data. For this, we postulate that the common signal observed by PTAs (and consistent with the increased evidence recently reported) is an astrophysical background sourced by a single MBHB population. 
We then constrain the LISA detection rate, $\mathcal{R}$, in the mass-redshift space by combining our Bayesian-inferred merger rate with LISA’s sensitivity to spin-aligned, inspiral-merger-ringdown waveforms. 
Using an astrophysically-informed formation model, we predict a $95\%$ upper limit on the detection rate of $\mathcal{R} <
134\,{\rm yr}^{-1}$ 
for binaries with total masses in the range $10^7 - 10^8\,\Msol$. 
For higher masses, i.e., $>10^8\,\Msol$, we find $\mathcal{R} < 2\,(1)\,\mathrm{yr}^{-1}$ using an astrophysically-informed (agnostic) formation model, rising to $11\,(6)\,\mathrm{yr}^{-1}$ if the LISA sensitivity bandwidth extends down to $10^{-5}$ Hz.
Forecasts of LISA science potential with PTA background measurements should improve as PTAs continue their search.
\end{abstract}


\section{Introduction}
\label{sec:Intro}

The formation and evolutionary paths of black holes observed at the centers of galaxies are fundamental open problems in astrophysics. While black holes with masses $\sim 10^9\,\Msol$ are likely already present at redshift $z \gtrsim 7.5$~\citep{2021ApJ...907L...1W} and are essentially ubiquitous in the cores of galaxies in the local Universe \citep{Kormendy:1995,Kormendy:2013,Heckman:2014}, the details of how they form, evolve, and interact with their host galaxies are still largely unclear. 

The mergers of binaries composed of comparable mass black-holes with total mass $M \gtrsim 10^6\,\Msol$ are a prime source for gravitational-wave (GW) detectors to probe astrophysical and cosmological uncertainties across cosmic time \citep{Sathyaprakash:2009,BailesEtAl:2021,AuclairEtAl:2022,Amaro-Seoane:2023}. 

Two observational windows of GWs allow us to study these massive black-hole binaries (MBHBs): the ultra-low ($\sim 1\,\mathrm{nHz} - 1\,\mu\mathrm{Hz}$) and low ($\sim 0.1\,\mathrm{mHz} - 100\,\mathrm{mHz}$) frequency regimes, the focus of pulsar timing arrays (PTAs)~\citep{FosterBacker:1990} and 
Laser Interferometer Space Antenna (LISA)~\citep{Amaro-Seoane:2017} observations, respectively. 
Theoretical modelling of sources of interest for PTAs and LISA have generally proceeded separately, as PTA observations are mainly sensitive to higher mass ($M \sim 10^8$--$10^{10}\,\Msol$) binaries at low-to-moderate redshift ($z \lesssim 2$) whereas LISA will provide information mainly about lighter ($M \sim 10^5$--$10^6\,\Msol$) binaries at high redshift ($z \approx 1$--$10$ and beyond). 

While the peak sensitivities of the two observatories are in mostly different portions of the mass-redshift parameter space, they still overlap and can be complementary~\citep[see also, e.g.'s,][]{SesanaEtAl:2008, Spallicci2013,EllisEtAl:2023}. More specifically, if one assumes that there is a (dominant) cosmic population of MBHBs spanning the full mass range $\sim 10^6$--$10^{10}\,\Msol$ PTAs and LISA will \textit{jointly} provide the tightest constraints on its properties. This is particularly timely as PTAs may observe a stochastic GW background (SGWB) before LISA is in science operation (2034+). In other words, direct observational results on MBHBs obtainable in the next few years can be used to make falsifiable predictions once LISA is in orbit.

To illustrate our point, we consider the recent results from the pulsar timing array collaborations, the North American Nanohertz Observatory for Gravitational waves~\citep[NANOGrav;][]{NANOGrav12p5yr:2020}, the Parkes PTA~\citep[PPTA;][]{PPTA:2021}, and the European PTA~\citep[EPTA;][]{EPTA:2021}, all of which combine the data within the umbrella of the International PTA~\citep[IPTA;][]{IPTA:2022}. They have each identified a statistically consistent common red-stochastic signal in the timing residuals of the pulsars constituting their arrays of unknown origin. We assume, purely for the sake of demonstration, that this signal is produced by a SGWB generated by a cosmic population of MBHBs described by some underlying (phenomenological) model. We show that the properties of the MBHB population inferred from the PTA results directly translate into predictions for the number of MBHB mergers -- and their properties, \textit{i.e.} masses and redshift -- that LISA will observe. 

More recently (while this work was under review), the PTA groups have announced increased evidence that the observed signal has a GW origin. 
The reported significance of a GW origin is between $2\sigma$ and $4\sigma$ from the EPTA~\citep{ETPAGWBAntoniadisEtAl:2023}, PPTA~\citep{PPTAGWReardonEtAl:2023}, NANOGrav~\citep{NANOGravGWBAgazieEtAl:2023}, and the Chinese PTA~\citep[CPTA,][]{CPTAGWBXuEtAl:2023}.
The nature of the signal is uncertain and various sources are being investigated including MBHBs, dark matter and the early Universe~\citep[e.g.][]{EPTAImplications:2023,NANOGravImplicationsMBHB:2023,NANOGravImplicationsNewPhysics:2023}.
The analysis presented in this paper is based on results from the IPTA second data release~\citep{IPTA:2022} which are consistent with the most recent PTA announcements. 

Theoretical estimates of the SGWB amplitude in the PTA band are uncertain~\citep[e.g.,][]{Sesana:2013, KelleyEtAl:2017b, ChenYuAndLu:2020, SykesEtAl:2022}. Likewise, the merger rate of MBHBs based on galactic evolution models are uncertain, mainly due to incomplete knowledge of galactic formation at high redshift \citep{Sesana:2021}. LISA is generally expected to observe between $\sim 1 - 100$ MBHBs per year~\citep{Rhook:2005,Sesana:2011,Klein:2016,Katz2020,Barausse:2020}. These merger rates are dominated by binaries in the mass range $M \sim 10^6 - 10^7\,\Msol$ independent of the uncertainties of the possible formation scenarios~\citep{Bonetti:2019}, implying it may be challenging for LISA to observe binaries with higher mass, i.e., $M \gtrsim 10^8\,\Msol$. 

LISA's ability to detect such black holes depends on the physical attributes of the detector itself, such as the detector's lower frequency limit \citep{Katz:2019}, but also on waveform modelling assumptions when computing the signal-to-noise ratio (SNR) of LISA. For example, higher order multipole modes can enter the LISA detection band at higher frequencies and extend the duration of a very massive binary signal in the LISA band to increase its accumulated SNR. 
Although this increase is at-best modest, it can make or break the detection of higher mass binaries that reside on the edge of LISA detectability. Importantly, using our framework for PTA-constrained merger rate estimates, we find that the LISA \emph{detection rate} of higher mass binaries, i.e., $M \gtrsim 10^8\,\Msol$, is sensitive to these instrumental and modelling assumptions. 

This paper is organised as follows. In Section~\ref{sec:models} we introduce the phenomenological models we use to describe the merger rate of MBHBs. In Section~\ref{sec:PTA}, under the assumptions mentioned above, we use the PTA observations to place constraints on the MBHB population parameters for the two models used in~\citet{MiddEtAl:2021} and compute the \textit{merger} rate of sources of interest to LISA. In Section~\ref{sec:LISA}, we revisit the LISA sensitivity to MBHBs by including in the GW radiation modes higher than the dominant $\ell = |m| = 2$ and account for the (pessimistic) low-frequency limit at $0.1\,\mathrm{mHZ}$, which plays a particularly significant role; by combining these results with those of Section~\ref{sec:PTA} we compute the LISA \textit{detection} rate of these binaries. Finally, in Section~\ref{sec:PE}, we randomly draw a few binaries from the population and use a Bayesian analysis on the full time-delay-interferometry LISA observables to forecast the information that LISA will be able to gather from the observations of these systems. 
In Section~\ref{sec:Disc} we conclude and discuss implications of the results.

\section{Model for the merger rate of massive black hole binaries}
\label{sec:models}

We first briefly review the phenomenological models that we use to describe the merger rate of MBHBs.

Generically, these models represent the merger rate of the MBHB population with a function, ${\cal F}(\boldsymbol{\lambda})$ 
\citep{Phinney:2001},  
 \begin{equation}
    {\cal F}(\boldsymbol{\lambda}) \equiv 
    \frac{\mathrm{d}^3 N(\boldsymbol{\lambda})}
    {\mathrm{d}V_{\rm c} 
    \mathrm{d}t_\mathrm{r} 
    \mathrm{d}\logten \Mc}\,.
    \label{eqn:model}
\end{equation}
Here, $N$ is the number of MBHB mergers per unit co-moving volume, $V_{\rm c}$, (source-frame) time, $t_\mathrm{r}$, and logarithmic (source-frame) chirp mass $\Mc$, 
where $\Mc = (m_1 m_2)^{3/5} (m_1+m_2)^{-1/5}$ for a binary with individual (source-frame) mass components $m_{1,2}$.
The parameter vector $\boldsymbol{\lambda}$ in Eq.~(\ref{eqn:model}) specifies the hyper-parameters that describe the population. 
Different astrophysical assumptions necessarily provide different functional forms for Eq.~(\ref{eqn:model}). 

Here, we are ultimately interested in the number of mergers per unit observer time $t$ within a redshift-chirp mass shell between $z$ and $z+\mathrm{d}z$ and $\mathcal{M}$ and $\mathcal{M}+\mathrm{d}\mathcal{M}$, 
\begin{equation}
\frac{\mathrm{d}^3N(\boldsymbol{\lambda})}{\mathrm{d}t \mathrm{d}z \mathrm{d}\logten \mathcal{M}} =
\frac{\mathrm{d}^3N(\boldsymbol{\lambda})}
{\mathrm{d}V_c \mathrm{d}t_\mathrm{r} \mathrm{d}\logten \mathcal{M}}\frac{\mathrm{d}V_{\rm c}} {\mathrm{d}z} \frac{\mathrm{d}t_\mathrm{r}} {\mathrm{d}t}
\label{eqn:d3N_dtdzdM}
\end{equation}
where $\mathrm{d}V_c/\mathrm{d}z = 4\pi c D_{\rm L}^2/\Ho(1 + z)^2 E(z)$ is the differential co-moving volume \citep{Hogg:1999} (we assume sources are distributed uniformly in the Universe), 
$\mathrm{d}t_\mathrm{r}/\mathrm{d}t = (1 + z)^{-1}$, $D_{\rm L}$ is the luminosity distance between source and observer, $\Ho$ is the present-day Hubble parameter, $E(z) = ( \OmegaM (1 + z)^3 + \OmegaL )^{1/2}$ for a flat Universe, and $\OmegaM$ and $\OmegaL$ are the mass and $\Lambda$ density parameters, respectively, \citet{Hogg:1999}.
Integrating Eq.~(\ref{eqn:d3N_dtdzdM}) over the relevant redshift and mass intervals provides the merger rate $\dot{N} \equiv \mathrm{d}N/\mathrm{d}t$ as measured by an observer. 

We consider the two models used in~\cite{MiddEtAl:2021} for the population of MBHBs: the ``agnostic'' model which imposes minimal assumptions~\citep{MiddletonEtAl:2016}, and a more ``astrophysically-informed'' model which accounts for various observational and theoretical astrophysics inputs in the model design, see
~\citep{ChenSesanaConselice:2019}. 
For complete descriptions of the details of these two models, see~\citet{MiddletonEtAl:2016, MiddEtAl:2018, MiddEtAl:2021, ChenEtAl:2017, ChenSesanaDelPozzo:2017, ChenSesanaConselice:2019}.
\begin{figure*}
	\centering
    \includegraphics[width=\textwidth]{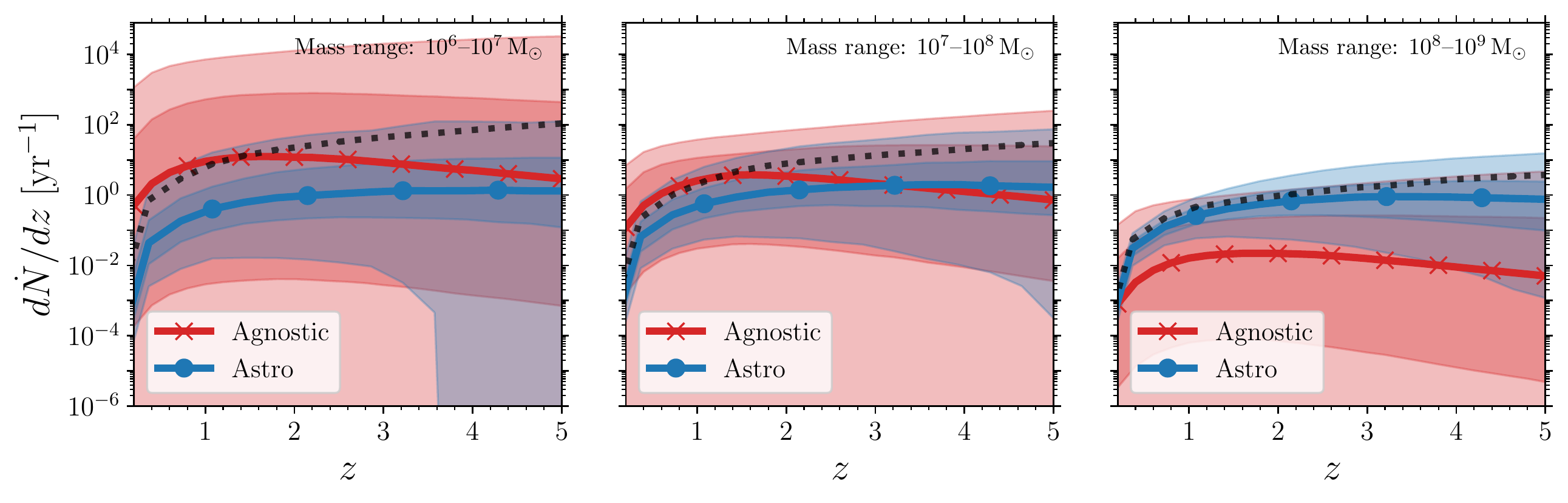}
    \caption{\label{fig:IPTAdNdzCombined}
    Posterior distributions of the MBHB merger rate per year per unit redshift over the range $z = 0 - 5$. The three panels correspond to three total mass ranges, $M = 10^6$--$10^7\,\Msol$, $10^7$--$10^8\,\Msol$, and $10^8$--$10^9\,\Msol$ for the left, middle, and right panels, respectively. In each panel, data from the agnostic (astro-informed) formation model is shown in red (blue), where the solid line marked by crosses (circles) is the median and the dark and light shaded regions represent the central $50\%$ and $90\%$ credible regions, respectively. These merger rates are inferred using the IPTA DR2 results. The black dotted line is the 99.5 percentile of the prior distribution for the astro-informed model.
    }
\end{figure*}

In the agnostic model,  Eq.~(\ref{eqn:model}) takes the simple parametric form:
\begin{equation}
    \frac{\mathrm{d}^3 N(\boldsymbol{\lambda})}{\mathrm{d} V_{\rm c} \mathrm{d} t_\mathrm{r} \mathrm{d} \logten \Mc} = 
    \ndot 
    \left( \frac{\Mc}{\Mscale} \right)^{-\alphaM}
    e^{-\Mc/\Mstar}
    \left(1+z\right)^{\beta} e ^{-z/z_0} \,.
    \label{eqn:model_agn}
\end{equation}
The model is characterised by five population hyper-parameters $\boldsymbol{\lambda} = \{\ndot, \alphaM, \Mstar, \beta, z_0\}$. 
Here $\ndot$ is the number density of mergers per %
unit rest-frame time and co-moving volume. 
The parameters $\alphaM$ and $\Mstar$ describe the slope and cut-off of the distribution of sources in $\Mc$, respectively. 
The parameters $\beta$ and $z_0$ provide the equivalent function for the distribution of sources in $z$. 
This model assumes that binaries merge in circular orbits driven by radiation reaction alone.
In~\cite{MiddEtAl:2021} it was assumed (somewhat arbitrarily) that this model is valid in the redshift range $0 \le z \le 5$ and in the chirp mass range $10^{6}\,\Msol \le \Mc \le 10^{11}\,\Msol$. 
In this work, we consider a total mass range of $10^6\,\Msol \le M \le 10^9\,\Msol$ (see Section~\ref{sec:PTA}). 
We therefore extrapolate at the low mass edge from $\Mc=10^6\,\Msol$ to $3.7\times10^5\,\Msol$ which corresponds to total mass $M = 10^6\,\Msol$ for a constant mass ratio $q=1/3$.
The model is agnostic in the sense that it allows for a wide range of distributions as broad prior ranges are used for each of the parameters (see Appendix~\ref{sec:iptaResults}). 

The astrophysically-informed model~\citep{ChenSesanaConselice:2019} is described by $18$ population hyper-parameters and allows for binaries to have a non-zero eccentricity. Of these parameters, $16$ are related to astrophysical observables which are informed by priors from observations and simulations. In brief, the number of MBHB mergers is linked to the number of galaxy mergers through a $M_{\rm gal}-M_{\rm BH}$ relation (three parameters). Galaxy mergers are described by the galaxy stellar mass function (five parameters), the fraction of galaxies in pairs (four parameters), which are then assumed to merge within a given time (four parameters). The final two parameters are related to the effects of the environment in which binaries evolve: the eccentricity and a parameter that depends on the stellar density in the galactic core 
and describes the interaction of a binary with the environment. 
The mass range for this model is determined by the galaxy masses $10^9$--$10^{12}\,\Msol$, which translates to $\approx 10^{6.3}$--$10^{9.3}\,\Msol$ in $\Mc$. Like the agnostic model, we extrapolate at the low mass end to $M=10^6\,\Msol$. 
In redshift, the astrophysically-informed model 
assumes most PTA-sensitive sources are at low redshift $z \leq 1.5$. Here we consider it valid up to redshift, $z=5$ as was done in~\cite{MiddEtAl:2021}. 

In both of our MBHB formation models, we assume cosmological parameters 
$H_0 = 70$ 
$\mathrm{km}^{-1}\mathrm{s}^{-1}\mathrm{Mpc}^{-1}$, $\OmegaM=0.3$, and $\OmegaL=0.7$, 
which are consistent with those from the most recent Planck cosmology~\citep{Planck:2018}.

\section{Black hole binary population constraints from pulsar timing arrays}
\label{sec:PTA}

The incoherent superposition of radiation from the %
cosmic MBHB population produces an isotropic, Gaussian, unpolarized SGWB with a characteristic amplitude~\citep{Phinney:2001} 
at GW frequency $f$, 
\begin{eqnarray}
h_{\rm c}^2 (f) & = & \frac{4 G^{5/3}}{3\pi^{1/3}c^2}f^{-4/3} \times
\nonumber \\
&&
          \int \mathrm{d} \Mc 
          \int \mathrm{d} z \left( 1+z \right)^{-1/3} 
          \Mc^{5/3}
          \frac{\mathrm{d}^3N(\boldsymbol{\lambda})}{\mathrm{d}z \mathrm{d}V_c \mathrm{d}\Mc} \,,
          \label{eq:hc_f}
\end{eqnarray} 
where $G$ and $c$ are the gravitational constant and speed of light, respectively.
In principle, the rate ${d^3N(\boldsymbol{\lambda})}/{dz dV_c d\mathcal{M}}$ can be
found from Eq.~(\ref{eqn:model}) in a similar manner as was done for Eq.~(\ref{eqn:d3N_dtdzdM}).

Current PTAs are most sensitive to a SGWB over a small frequency interval spanning the few lowest possible frequency bins associated with the period covered by the observations, 
which at present is $\approx 20\,\mathrm{yr}$, see \textit{e.g.}~\cite{NANOGrav12p5yr:2020, PPTA:2021, EPTA:2021, IPTA:2022}. This implies that %
the SGWB characteristic amplitude, regardless of its physical origin, is well described over this sensitivity range by a power-law, %
\begin{equation}
	h_{\rm c}(f) = \Ayr \left( \frac{f}{1{\rm yr}^{-1}} \right)^{\alpha},
         \label{eq:hc_power_law}
\end{equation}
where $\Ayr$ is the unknown SGWB amplitude at GW frequency $f$ of $1\,{\rm yr}^{-1}$ and $\alpha$ is the spectral index. 
For a background produced by MBHBs $\alpha = -2/3$, cfr. Eq.~(\ref{eq:hc_f}).

The most recent observational results from the PTA consortia report statistically consistent evidence of a common stochastic signal in the pulsar timing residuals~\citep{NANOGrav12p5yr:2020, PPTA:2021, EPTA:2021, IPTA:2022}. As we have already stressed in Section~\ref{sec:Intro}, further work and observations are required to ascertain the origin of the signal. However, if we assume it to be generated by a SGWB, the corresponding median values of $A_{\rm 1yr}$ reported by each of the PTAs are in the range $1.92 \text{--} 2.95 \times 10^{-15}$. 

To illustrate our main contention that PTAs operating now can inform future LISA observations, we shall assume that this signal is %
due to a SGWB %
whose origin is a single cosmic population of MBHBs. 
We consider the IPTA DR2 results~\citep{IPTA:2022} to infer constraints on the population hyper-parameters of the models described in Section~\ref{sec:models} from which we compute the posterior probability distributions on the MBHB merger rate. That is, with the PTA data $d$ we evaluate,
\begin{equation}
	p(\boldsymbol{\lambda} | d) \propto  {\cal L}(d | \boldsymbol{\lambda})\, p(\boldsymbol{\lambda})\,,
\end{equation}
where $p(\boldsymbol{\lambda})$ are the priors on the population parameters, and the likelihood ${\cal L}(d | \boldsymbol{\lambda})$ is computed from the %
IPTA DR2 free-spectrum-analysis posteriors, which are converted into characteristic amplitude $h_c$ for $\alpha = -2/3$,
by taking the 
lowest five frequency bins 
and summing the log-likelihoods at those values, %
as described in the Methods in~\cite{2021NatAs...5.1268M}. 
The agnostic model assumes circular binaries, and therefore requires $h_c$ from a single frequency, which we choose to be $1\,{\rm yr}^{-1}$.
We assume priors $p(\boldsymbol{\lambda})$ as those considered in~\cite{MiddEtAl:2021} for both formation models. Not surprisingly, the posterior distributions on $\boldsymbol{\lambda}$ we obtain with IPTA results are very similar to those reported in ~\cite{MiddEtAl:2021}, which were computed using the NANOGrav 12.5 year results.
as the IPTA and NANOGrav results are statistically consistent. 
We use \textsc{cpnest}~\citep[a nested sampling implementation][]{VeitchVecchioCPNEST:2010,cpnest} and \textsc{ptmcmc}~\citep[a Markov Chain Monte Carlo implementation][]{ptmcmc} 
for the sampling. 
For completeness we include the full posteriors in Appendix~\ref{sec:iptaResults}.

\begin{figure}
    \includegraphics[width=0.49\textwidth]{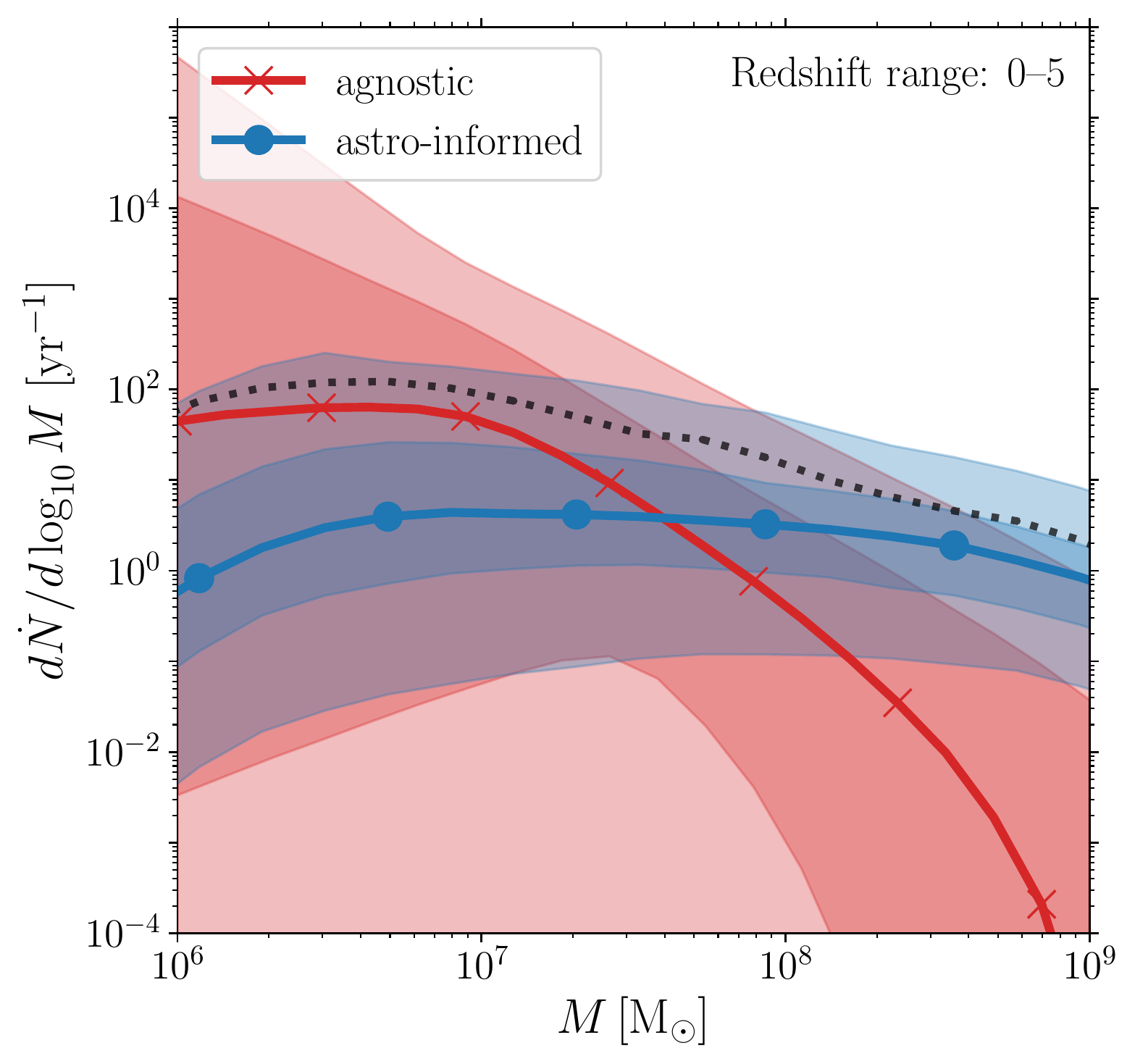}
    \caption{\label{fig:IPTAdNdlogMCombined}
    Posterior distributions of %
    the MBHB merger rate per year per unit logarithmic total-mass %
    over the range $M = 10^6 - 10^9\,\Msol$. 
    As in Fig.~\ref{fig:IPTAdNdzCombined}, the agnostic 
    (astro-informed) model is shown in red (blue), 
    where the solid line marked by crosses (circles) is the median and the dark and light shaded regions represent the central $50\%$ and $90\%$ credible regions, respectively. These merger rates are inferred using the IPTA DR2 results.
    Sources are integrated over a redshift range of $0$ to $5$. The black dotted line is the 99.5 percentile of the prior distribution for the astro-informed model.
    }
\end{figure}

For a given formation model, the above procedure provides a posterior distribution on the MBHB merger rate, $\mathrm{d}^3 N(\boldsymbol{\lambda})/\mathrm{d} t \mathrm{d} z \mathrm{d} \logten \mathcal{M}$. Using Eq.~\ref{eqn:d3N_dtdzdM} we convert the intrinsic merger rate to rate in \textit{observer} time and integrate over $\Mc$ or $z$ to derive the posterior distribution on the merger rate as a function of $z$, as shown in Fig.~\ref{fig:IPTAdNdzCombined}, or merger rate as a function of (source-frame) total mass, as shown in Fig.~\ref{fig:IPTAdNdlogMCombined}. In both figures, the posteriors of the agnostic (astro-informed) model are shown in red (blue).
provides a set of posterior samples, i.e., $\mathrm{d}^3N(\boldsymbol{\lambda})/ \mathrm{d} t \mathrm{d} z \mathrm{d} \logten \mathcal{M}$, that correspond to a MBHB population distribution. 
With this distribution, we then either integrate over $\Mc$ and $t_{\rm obs}$ (mission duration) to estimate the number of mergers anticipated by LISA as a function of $z$, as shown in Fig.~\ref{fig:IPTAdNdzCombined}, or we integrate over $z$ and $t_{\rm obs}$ to estimate the number of mergers as a function of $\Mc$, as shown in Fig.~\ref{fig:IPTAdNdlogMCombined}. 
In both figures, the posteriors of the agnostic (astro-informed) model are shown in red (blue). %

The left, middle, and right panels in Fig.~\ref{fig:IPTAdNdzCombined} correspond to ranges of total mass $M = 10^{6}-10^{7}\,\Msol$, $10^{7}-10^{8}\,\Msol$, and $10^{8}-10^{9}\,\Msol$, respectively, where we have assumed a constant mass ratio $q=1/3$ to convert from $\Mc$ to $M$.
The lowest mass range (left panel) produces the largest numbers of mergers, consistent with previous analysis
\cite{MiddEtAl:2021}.
While the agnostic model predicts $< 1$ high-mass ($10^{8}-10^{9}\,\Msol$) merger per year per unit redshift, displayed in the right panel, the astro-informed model affords $\gtrsim 1\,\mathrm{yr}^{-1}$ such mergers. 
The explanation for these differences lies in the priors for each model. 
The priors of the astro-informed model, shown by the black dotted line,  
are set by additional observational/theoretical astrophysics considerations which are not included in the agnostic model (the full set of priors on each individual parameter are are shown by the green lines and contours in Fig.~\ref{fig:astroModelIPTA} in Appendix~\ref{sec:iptaResults}).
As a consequence they allow narrower ranges of $\dot{N}$ distributions compared to the agnostic model which uses uniform and uninformative priors over the parameters.

Addditionally, the agnostic model permits realisations with a high number of low-mass binaries around $M\gtrapprox 10^6\,\Msol$ and correspondingly fewer high-mass mergers, scenarios that are disallowed by the priors of the astro-informed model~\citep[see e.g., Fig.~3 of][]{MiddEtAl:2021}. 
This key difference between the two models is more apparent in Fig.~\ref{fig:IPTAdNdlogMCombined}. The median of the posteriors of the agnostic model vanishes at large $M$ before the median of the astro-informed model's posteriors, which are generally %
flatter 
across this mass range, due to the astro-informed model's prior distribution indicated by the black dotted line. 
This highlights how the merger rates calculated here are sensitive to population modelling assumptions, in part due to the weak constraints from PTA results.

Now that we have established the PTA-constrained estimates for the merger rates of MBHBs, we turn next to the detectability of MBHBs with LISA to ultimately estimate their detection rates.

\begin{figure*}
\begin{center}
\includegraphics[width=1\textwidth]{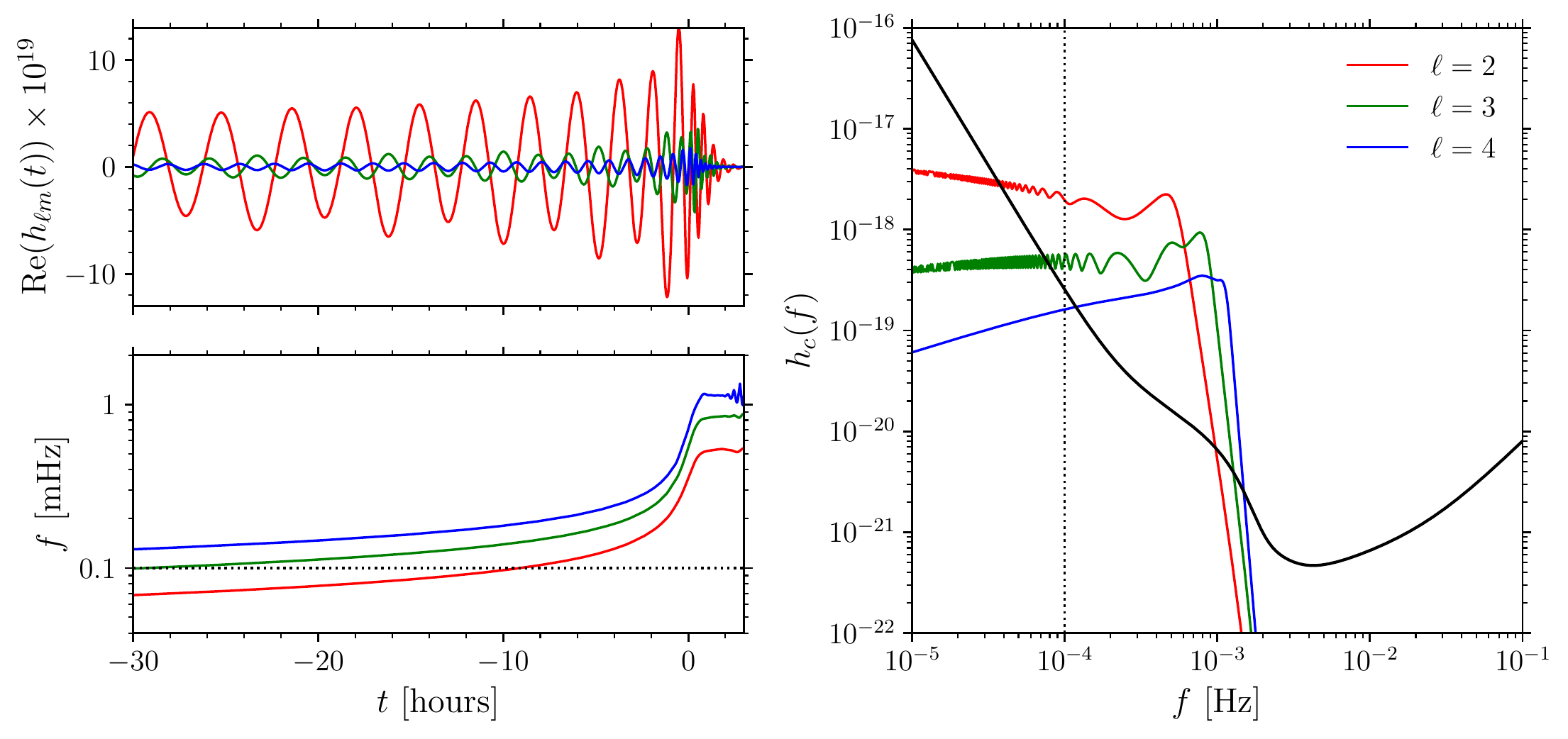}
\end{center}
\caption{
    The real part of the strain amplitude $h_{\ell m}(t)$ (top-left panel) and the frequency $f$ (bottom-left panel) of gravitational signals composed of the $\ell = 2,~3$, and 4 modes which correspond to the red, green, and blue solid lines, respectively. The characteristic strain $h_c(f)$ of each signal (right panel) passes through the LISA detection band, i.e., the characteristic noise amplitude of the noise power spectral density $h_n(f)$, shown by the black solid line. The dotted black line is the usual lower frequency limit of LISA, $f_{\rm low} = 10^{-4}$ Hz. Each of the three signals assume the same binary mass ratio $q = 1/3$, source-frame total mass $M = \num{3e7}\,\Msol$, luminosity distance $10^5$ Mpc, and inclination $\iota = \pi/2$.
    }\label{fig:StrainFreqPSD}
\end{figure*}

\section{LISA detections of black hole mergers}
\label{sec:LISA}

While PTAs detect (primarily) the contribution from the ensemble of the MBHB population, LISA will resolve the coalescences of individual MBHBs from the population. Despite naively appearing to probe separate domains, these detectors are complimentary as constraints from one can inform our expectations for the other. In this section, we demonstrate how combining PTAs and LISA can provide useful astrophysical insight. First, let's examine the waveforms of MBHBs and LISA's detection capabilities.

The strain produced by a MBHB can be represented as a multipole expansion with basis functions $_{-2}Y_{\ell m}(\theta,\phi)$, the spin-weighted spherical harmonics of spin weight $-2$, and with coefficients $h_{\ell m}(t)$, 
\begin{equation}
h(t) = h_+ (t) - i h_\times (t) = \sum_{\ell \ge 2} \sum_{m = -\ell}^{\ell} h_{\ell m}(t)\, _{-2}Y_{\ell m}(\theta,\phi)\,,
\label{eqn:Strain}
\end{equation}
where $h_{+, \times}(t)$ are the polarisation amplitudes. Throughout this work, we use the waveform approximant 
\textsc{IMRPhenomXHM}~\citep{PrattenEtAl:2020,GarciaQuirosEtAl:2020} to compute Eq.~(\ref{eqn:Strain}): it describes the full inspiral-merger-ringdown radiation produced by the coalescence of binary systems in which the spins of the black holes are aligned to the orbital angular momentum of the binary. The multipoles $h_{\ell m}$ are calibrated to numerical relativity for mass ratios $q > 1/18$, and the approximant includes the $(\ell,|m|) = \lbrace (2,2),(2,1),(3,3),(3,2),(4,4)\rbrace$ modes. 

A single multipole mode $(\ell,\,m)$ is itself a complex-valued time series, $h_{\ell m}(t)$, which can be expressed via an amplitude $A_{\ell m}$ and phase $\Phi_{\ell m}$, 
\begin{equation}
h_{\ell m}(t) = h_+ (t) - i h_\times (t) = A_{\ell m}(t)e^{i\Phi_{\ell m}(t)}\,. \notag
\end{equation}
The corresponding frequency of this mode is the rate of change of the phase angle, 
\begin{equation}
f_{\ell m}(t) = \frac{1}{2\pi} \frac{\mathrm{d}}{\mathrm{d}t} \Phi_{\ell m}(t) \,.
\end{equation}
The top-left (bottom-left) panel of Fig.~\ref{fig:StrainFreqPSD} displays the amplitude Re$(h_{\ell m}(t))$ (frequency $f_{\ell m}(t)$) for three waveforms composed of $\ell = 2$, $\ell = 3$, and $\ell = 4$ modes that correspond to the red, green, and blue solid lines. Although the leading $\ell = 2$ modes dominate in amplitude Re$(h_{\ell m}(t))$, the frequency $f_{\ell m}$ is higher for the subdominant $\ell = 3$ and 4 modes.

The time spent by a GW signal in the detection band of a detector that is sensitive to frequencies above $f_{\rm low}$ is approximately,
\begin{equation}\label{eqn:Tau}
\tau = \frac{5}{256} \left(\frac{G \mathcal{M}}{c^3}\right)^{-5/3} \left(\pi f_{\rm low}\right)^{-8/3} \left(\frac{\ell}{2}\right)^{8/3}\,.
\end{equation}
For high-mass black-hole binaries and a given $f_{\rm low}$, the leading mode $\ell = m = 2$ spends less time in the LISA detection band compared to the higher modes $\ell = m = 3,~4$. This implies that the higher modes can improve the detectability of higher mass binaries despite radiating more quietly than the lower modes.

\begin{figure*}
\begin{center}
\includegraphics[width=1\textwidth]{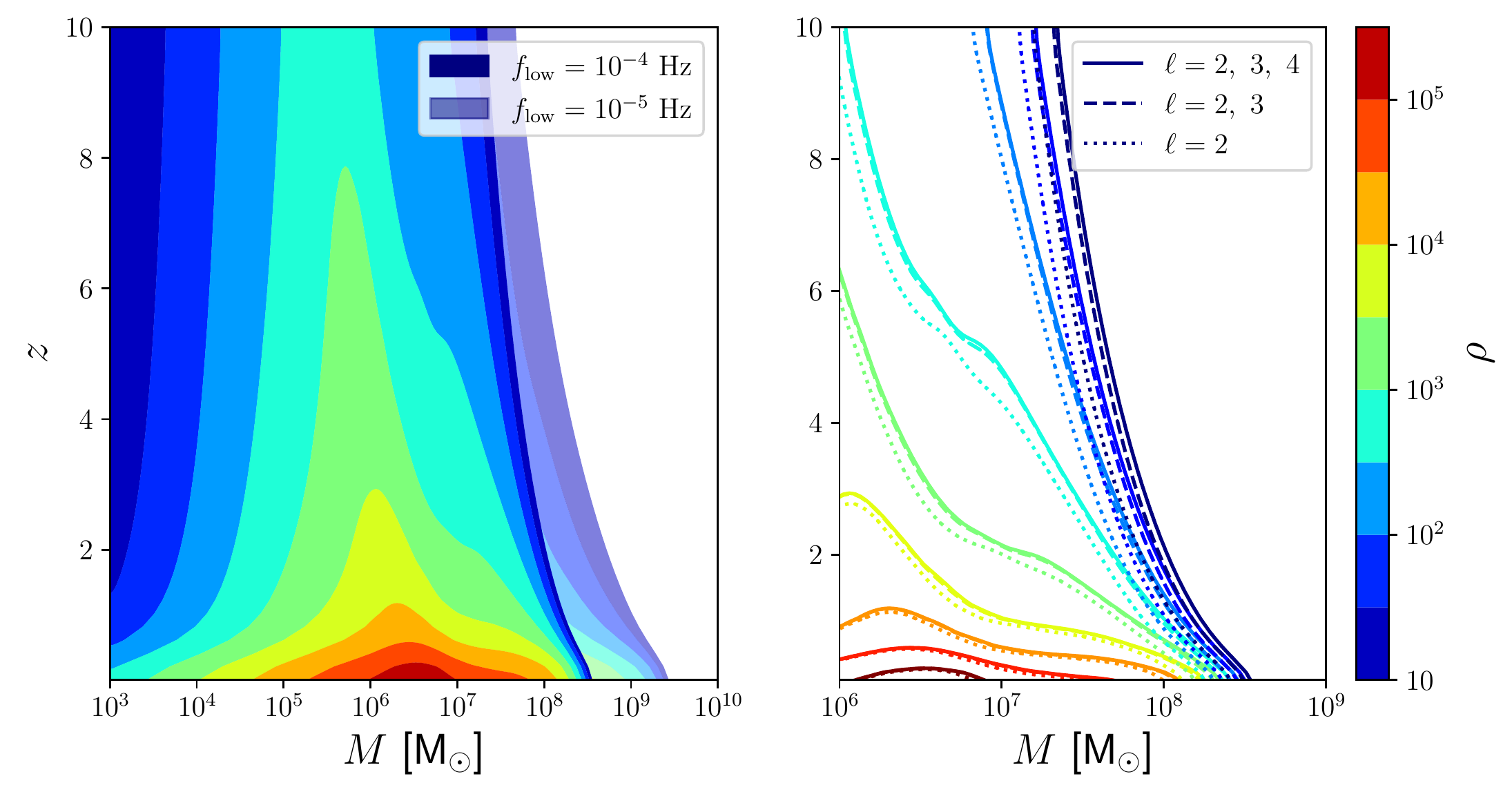}
\end{center}
\caption{\label{fig:Waterfall}
    Contours of LISA signal-to-noise versus redshift $z$ and source-frame total binary mass $M$ assuming a mass ratio $q = 1/3$, zero spins, inclination $\cos\iota = 0.8$, and flat FLRW cosmology with cosmological parameters taken from \citet{Planck:2018}. The completely (transparently) filled contours in the left panel assume a lower frequency cut off $f_{\rm low} = 10^{-4}$ Hz ($f_{\rm low} = 10^{-5}$ Hz) and are computed with the available $\ell = 2$, $\ell = 3$, and $\ell = 4$ modes of the \textsc{IMRPhenomXHM} waveform approximant. The right panel shows a zoomed-in portion of these signal-to-noise contours as solid lines, and the dashed and dotted lines are computed with only the $\ell = 2$ and $\ell = 3$ modes and only the $\ell = 2$ modes, respectively. 
    }
\end{figure*}

The characteristic strain of a signal, useful when considering the response of a GW detector, is defined as $[h_c(f)]^2 = 4f^2|\tilde{h}(f)|^2$ \citep{Moore:2015} where $\tilde{h}(f)$ is the strain in the frequency domain, i.e., $\tilde{h}(f) = \int_{-\infty}^\infty h(t)\exp[-2\pi ift]\mathrm{d}t$. Analogously, $[h_n(f)]^2 = fS_n(f)$ gives the amplitude that describes the noise of a detector where $S_n(f)$ is the (one-sided) power spectral density of the noise. The right panel of Fig.~\ref{fig:StrainFreqPSD} shows $h_c(f)$ for the same three sets of modes where the black solid line is the noise amplitude $h_n(f)$ of the LISA detector assuming $S_n(f)$ is composed of instrumental \citep{BabakEtAl:2021} and galactic confusion \citep{BabakEtAl:2017} noises. The oscillations in $\ell = 2$ and 3 are due to contributions of the (2,1)+(2,2) and (3,2)+(3,3) subbands, respectively. Consistent with the bottom left panel, the right panel of Fig.~\ref{fig:StrainFreqPSD} demonstrates how higher modes enter the LISA detection band at higher frequencies and extends the duration of the signal from a high-mass binary in the LISA band. The low-frequency limit of the LISA detector, shown by the dotted black line, plays a similar role.

The presence of radiation multipoles higher than the dominant $\ell = |m| = 2$ mode is particularly important to take into account when considering observations of MBHBs whose total mass-redshift combination produce a signal close to the low-frequency sensitivity limit of LISA. These higher modes provide sensitivity to a larger portion of the mass-redshift parameter space than only considering the $\ell = |m| = 2$ mode such as in Fig.~3 of \citet{Amaro-Seoane:2017}. Motivated by this, next we apply the above methodology to explore the detectability of MBHB mergers with LISA and the impact of higher modes.

For an L-shaped GW detector, the SNR is \citep{Moore:2015},
\begin{equation}\label{eqn:SNR}
\rho^2 = \langle h | h \rangle = \int_{-\infty}^{\infty} \left[ \frac{h_c(f)}{h_n(f)} \right]^2 \mathrm{d}(\log f)\,,
\end{equation} 
where $\langle h | h \rangle$ denotes the noise-weighted inner product \citep{Cutler:1994}. 
The SNR for a LISA-like detector can, in the limit that GW wavelengths are much larger than the LISA arm length, be approximated as the sum of the SNRs of two L-shaped detectors, each given by Eq.~(\ref{eqn:SNR}) above, %
and accounting for the $\pi/3$ rather than $\pi/2$ angle between the LISA arms, i.e., LISA's SNR is 
$\rho^2 = \langle H_1 | H_1 \rangle + \langle H_2 | H_2 \rangle$ 
where $H_i = \sqrt{3} (F_{i,+}h_+ + F_{i,\times}h_\times)/2$ are the responses for two sets of L-shaped arms assuming each measures the same MBHB signal with the same detector noise as in Fig.~\ref{fig:StrainFreqPSD} and $F_{i,+\times}$ are the beam-pattern coefficients \citep{Barack:2004}. We vary the MBHB total mass and redshift, and rather than marginalizing over the source spins, orientation, and location, here we assume that the MBHBs are nonspinning, oriented with modest inclination ($\cos\iota = 0.8$), directly above the detector with constant azimuth (i.e., $\theta_s = 0, \phi_s = \pi/4$) in the frame that is corotating with the detector, and that the GW polarization is constant (i.e., $\psi = \pi/6$).

The contours of constant SNR in the left panel of Fig.~\ref{fig:Waterfall} are computed as functions of the redshift $z$ and source-frame total mass $M$ and with waveforms [Eq.~(\ref{eqn:Strain})] composed of all the available multipole modes of \textsc{IMRPhenomXHM}, i.e., $(\ell \geq 2, |m| \leq \ell)$. As lower mass binaries produce longer lived signals in the LISA detection band, i.e., as shown by the black solid line in the right panel of Fig.~\ref{fig:StrainFreqPSD}, much of their accumulated SNR comes from the binary inspiral phase. 
Higher mass binaries spend less of their inspiral in the LISA band, i.e., see Eq.~(\ref{eqn:Tau}), and 
for binaries with $M \gtrsim 10^8\,\Msol$ essentially only the merger and ringdown are observable. Such signals are on the cusp of the low-frequency limit $f_{\rm low}$ of LISA, implying that the value of $f_{\rm low}$ is important for detecting very massive binaries. We demonstrate this with two values of $f_{\rm low}$ in the left panel of Fig.~\ref{fig:Waterfall}, where the solid-filled contours are computed assuming $f_{\rm low} = 10^{-4}$ Hz, and the faded contours are computed with a smaller frequency cut off $f_{\rm low} = 10^{-5}$ Hz. 
The smaller $f_{\rm low}$ allows binaries with mass $M \lesssim \num{5e8}\,\Msol$ and redshift $z \lesssim 2$ to have significantly larger SNR than compared to the higher $f_{\rm low}$, as more of the inspiral extends into the LISA window. Binaries with even higher mass are undetectable with the limit $f_{\rm low} = 10^{-4}$~Hz, and only become detectable with smaller $f_{\rm low}$ because these signals are already very short-lived in the peak of the LISA sensitivity window. This effect indicates that utilizing $f_{\rm low} \leq 0.1$ mHz 
will help to maximize the detection horizon, and hence the constraining power, of the LISA mission.

\begin{table*}
\caption{\label{tab:DRSummary}
The 5$^{\rm th}$, 50$^{\rm th}$, and 95$^{\rm th}$ percentiles of the detection rates (per year) $\mathcal{R}$ and the corresponding detectable fraction $F_{\rm det}$ of populations of MBHBs with LISA for the three mass bins: low $M = 10^6 - 10^7\,\Msol$, mid $M = 10^7 - 10^8\,\Msol$, and high $M = 10^8 - 10^9\,\Msol$. Four combinations of multipole modes (i.e., $\ell = 2$ or $\ell = 2,~3,~4$) and the LISA lower-frequency limit (i.e., $ 10^{-4}$ Hz or $ 10^{-5}$ Hz) are considered. The detection rates without (with) parenthesis are computed with merger rates from the agnostic (astro-informed) model of Section~\ref{sec:models}. An entry of 0 indicates that the rate is $< 0.1~{\rm yr^{-1}}$.
}
\vspace{0.1cm}
\def\arraystretch{1.2}  
\setlength{\tabcolsep}{0.7em} 
\centering
\begin{tabular}{c|cccc|cccc|cccc|}
\cline{2-13}
  \multicolumn{1}{c|}{} & \multicolumn{4}{c|}{$M = 10^6 - 10^7\,\Msol$} & \multicolumn{4}{c|}{$M = 10^7 - 10^8\,\Msol$} & \multicolumn{4}{c|}{$M = 10^8 - 10^9\,\Msol$} \\ 
  & $\mathcal{R}_5$ & $\mathcal{R}_{50}$ & $\mathcal{R}_{95}$ & $F_{\rm det}$ & $\mathcal{R}_5$ & $\mathcal{R}_{50}$ & $\mathcal{R}_{95}$ & $F_{\rm det}$ & $\mathcal{R}_5$ & $\mathcal{R}_{50}$ & $\mathcal{R}_{95}$ & $F_{\rm det}$ \\
   & [yr$^{-1}$] & [yr$^{-1}$] & [yr$^{-1}$] & & [yr$^{-1}$] & [yr$^{-1}$] & [yr$^{-1}$] & & [yr$^{-1}$] & [yr$^{-1}$] & [yr$^{-1}$] & \\
\hline
    \multicolumn{1}{|c|}{$\ell = 2$ \& $10^{-4}\,{\rm Hz}$} & 0 (0.1) & 75 (5) & 82700 (391) & 1 (1) & 0 (0.1) & 13 (4) & 373 (101) & 0.8 (0.7) & 0 (0) & 0 (0) & 1 (0) & 0.1 (0.0) \\
    \multicolumn{1}{|c|}{$\ell = 2$ \& $10^{-5}\,{\rm Hz}$} & 0 (0.1) & 75 (5) & 82700 (391) & 1 (1) & 0 (0.2) & 16 (7) & 464 (147) & 0.9 (0.9) & 0 (0.1) & 0.1 (0.8) & 5 (7) & 0.5 (0.3) \\
    \multicolumn{1}{|c|}{$\ell \geq 2$ \& $10^{-4}\,{\rm Hz}$} & 0 (0.1) & 75 (5) & 82700 (391) & 1 (1) & 0 (0.2) & 15 (6) & 434 (134) & 0.9 (0.9) & 0 (0) & 0 (0) & 2 (0) & 0.2 (0.0) \\
    \multicolumn{1}{|c|}{$\ell \geq 2$ \& $10^{-5}\,{\rm Hz}$} & 0 (0.1) & 75 (5) & 82700 (391) & 1 (1) & 0 (0.2) & 17 (7) & 477 (154) & 1 (1) & 0 (0.1) & 0.1 (1) & 6 (11) & 0.6 (0.4) \\
 \hline
\end{tabular}
\end{table*}

The right panel of Fig.~\ref{fig:Waterfall} shows how including higher modes can also modestly increase the SNR of MBHBs. The SNRs corresponding to the solid lines are identical to the solid-filled contours in the left panel, i.e., are computed with the $\ell = 2,~3,$ and 4 modes, and are monotonically larger than the corresponding SNR contours computed with the $\ell = 2$ and $\ell = 3$ modes (dashed lines), and with the $\ell = 2$ modes (dotted lines). This effect is essentially negligible for binaries with low mass ($M \lesssim 10^7$) and close proximity ($z \lesssim 2$), but is noticeable for binaries with higher total mass and distance. Again, this effect is most pronounced for very massive binaries whose short-lived signals (as seen by LISA) are extended with higher modes. We note that this effect is sensitive to the inclination of the source, i.e., it is largest for edge-on systems, and we choose a conservative inclination of $\cos\iota = 0.8$ in Fig.~\ref{fig:Waterfall}.

Both of the increases in SNR, due to smaller $f_{\rm low}$ or inclusion of higher modes, can make-or-break detections of higher-mass binaries, but the former is an instrumental feature of LISA while the latter is an observational modelling assumption. On this basis, we conclude that an optimal LISA detector 
should utilize as low of a frequency limit as possible; nevertheless, including higher modes in analysis of very massive mergers will systematically improve LISA's ability to probe these sources. As we shall see in Section~\ref{sec:PE}, including higher modes has the additional benefit of significantly improving the estimation of source parameters as it breaks the distance-inclination degeneracy. 

Now that we understand how LISA's sensivity depends on these various ingredients, we can estimate MBHB \emph{detection} rates for LISA by relating the MBHB \emph{merger} rates from PTAs in Section~\ref{sec:PTA} with the SNR of LISA. To produce Fig.~\ref{fig:Waterfall}, we computed SNRs in the long-wavelength approximation with a fitted curve for the detector noise. Instead, here we compute LISA SNRs using the full time-delay-interferometry implementation of \textsc{balrog}~\citep{Buscicchio:2021} (see Section~2 of~\cite{PrattenEtAl:2022} for a complete description) and the \textsc{IMRPhenomXHM} approximant~\citep{PrattenEtAl:2020,GarciaQuirosEtAl:2020} for the waveforms. 
We consider binaries with redshift $z = 0 - 5$ in three bins of (source-frame) total mass $M = 10^6 - 10^7\,\Msol$, $M = 10^7 - 10^8\,\Msol$, and $M = 10^8 - 10^9\,\Msol$ labeled low, mid, and high which correspond to the left, middle, and right panels of Fig.~\ref{fig:IPTAdNdzCombined}, respectively. The  total mass and redshift are randomly drawn from our PTA-constrained MBHB populations and the remaining extrinsic and intrinsic MBHB parameters are chosen randomly for each binary in the distribution.
We use $10,000$ draws for each $M$ range.

The detection rate $\mathcal{R}$ is the number of mergers with LISA SNR $\geq 12$ per year of observation time, i.e., $\mathcal{R} \equiv F_{\rm det}\dot{N}$ where $F_{\rm det}$ is the fraction of detectable mergers and $\dot{N}$ is the integrated merger rate from Eq.~(\ref{eqn:d3N_dtdzdM}) for the given mass-redshift bin. 
The merger rate can be sensitive to the formation model, as shown in Fig.'s~\ref{fig:IPTAdNdzCombined} and \ref{fig:IPTAdNdlogMCombined}, and can be as large as $\sim 10^4$ and $\sim 10^2$ for the agnostic and astro-informed models, respectively. Nevertheless, both models generate the most mergers in bin A and fewer mergers in bins B and C since the loud, very high-mass binaries that dominate the SGWB are outnumbered by the quieter low-mass binaries. Simultaneously, the SNR of LISA is largest for binaries in the low mass bin and decreases nearly monotonically  for higher mass binaries, as shown in Fig.~\ref{fig:Waterfall}. Therefore, one naively expects $\mathcal{R}$ to be largest for moderate-mass binaries (low bin) and smallest for very massive binaries (high bin) where modelling and instrumental assumptions will be important for MBHBs on the cusp of LISA detectability.

Our detection rates in these three mass-redshift bins are summarized in Table~\ref{tab:DRSummary} for both formation models and four combinations of multipole modes and the LISA low-frequency limit. 
We indeed find that $\mathcal{R}$ is generally largest in the low bin ($M = 10^6 - 10^7\,\Msol$), where the agnostic model predicts 
$\gtrsim 100$ times more detections compared to the astro-informed model at the 95$^{\rm th}$ percentile. 
The enormous SNR of LISA in this portion of the parameter space allows for all mergers to be detected, i.e., $F_{\rm det} = 1$, in either formation history. 

The detectability of MBHBs in the mid ($M = 10^7 - 10^8\,\Msol$) and high ($M = 10^8 - 10^9\,\Msol$) bins is more sensitive to the LISA low-frequency limit $f_{\rm low}$ and the inclusion of higher multipole modes. 
For the agnostic (astro-informed) formation model in the mid bin, $\mathcal{R}$ is $\approx20$\% ($\approx50$\%) larger with higher modes included and $f_{\rm low} = 10^{-5}$ Hz than with only the leading $\ell = 2$ modes and $f_{\rm low} = 10^{-4}$ Hz. 
Importantly, for the high bin, $\mathcal{R}$ is similarly boosted to 6 (11) $\mathrm{yr}^{-1}$ in the agnostic (astro-informed) formation model when higher modes and smaller $f_{\rm low}$ are assumed. The fraction of mergers that are detectable $F_{\rm det}$ in the high bin is larger for the agnostic model than the astro-informed model as the latter generates more mergers with higher mass, shown by Fig.~\ref{fig:IPTAdNdlogMCombined}. 
Thus, in our model, $\mathcal{R}$ is essentially only sensitive to the SNR cut-off in the high bin.  

In Table~\ref{tab:DRSummary} we use an entry of 0 to indicate $\mathcal{R} < 0.1~\mathrm{yr}^{-1}$. For the agnostic formation model, the 5$^{\rm th}$ percentile $\mathcal{R}_5$ is very small and we are only able to place upper limits on $\mathcal{R}$. 
However, the median $\mathcal{R}_{50} \approx 0.1~\mathrm{yr}^{-1}$ in the high bin, implying that a 10 $\mathrm{yr}$ LISA mission would detect at least 1 merger.
This is especially relevant for the astro-informed model, where $\mathcal{R}_5 \gtrapprox 0.1~\mathrm{yr}^{-1}$ in all three bins. 
We note that, in the first and third rows of Table~\ref{tab:DRSummary}, $\mathcal{R}$ is set to 0 for the astro-informed model in the high bin because $F_{\rm det} \sim 0.01$.

These results demonstrate that not only are PTA constraints of the SGWB capable of informing LISA detection rates, but that our framework can also probe the very uncertain formation of the MBHB population. We stress that higher modes and an optimistic value for $f_{\rm low}$ will aid the viability of such predictions.

\section{Parameter estimation with LISA}
\label{sec:PE}

\newcommand{\ASNRa}{276}
\newcommand{\ASNRb}{383}
\newcommand{\BSNRa}{7194}
\newcommand{\BSNRb}{7260}
\newcommand{\CSNRa}{745}
\newcommand{\CSNRb}{1068}
\newcommand{\DSNRa}{8}
\newcommand{\DSNRb}{78}
\newcommand{\ESNRa}{2269}
\newcommand{\ESNRb}{2595}



\newcommand{\APrimaryMassInj}{1.7}
\newcommand{\ASecondaryMassInj}{0.6}
\newcommand{\ARedshiftInj}{4.3}

\newcommand{\BPrimaryMassInj}{1.1}
\newcommand{\BSecondaryMassInj}{0.4}
\newcommand{\BRedshiftInj}{0.8}

\newcommand{\CPrimaryMassInj}{1.4}
\newcommand{\CSecondaryMassInj}{0.5}
\newcommand{\CRedshiftInj}{3.2}

\newcommand{\DPrimaryMassInj}{52.5}
\newcommand{\DSecondaryMassInj}{17.5}
\newcommand{\DRedshiftInj}{2.0}

\newcommand{\EPrimaryMassInj}{0.9}
\newcommand{\ESecondaryMassInj}{0.3}
\newcommand{\ERedshiftInj}{2.3}

\newcommand{\DimensionlessSpinOne}{0}
\newcommand{\DimensionlessSpinTwo}{0}
\newcommand{\EclipticLong}{2.0}
\newcommand{\sinEclipticLat}{0.3}
\newcommand{\InitialOrbitalPhase}{0.0}
\newcommand{\Polarization}{0.5}
\newcommand{\cosInclination}{0.8}
\newcommand{\MergerTime}{31536000}



\newcommand{\APrimaryMassPost}{1.73_{-0.02}^{+0.01}}
\newcommand{\ASecondaryMassPost}{0.574_{-0.005}^{+0.005}}
\newcommand{\ARedshiftPost}{4.30_{-0.04}^{+0.04}}
\newcommand{\AChiEffPost}{-0.001_{-0.005}^{+0.006}}
\newcommand{\AMergerTimePost}{0_{-9}^{+8}}
\newcommand{\ASkyAreaPost}{1.4}

\newcommand{\BPrimaryMassPost}{1.1252_{-0.0003}^{+0.0004}}
\newcommand{\BSecondaryMassPost}{0.3749_{-0.0001}^{+0.0001}}
\newcommand{\BRedshiftPost}{0.8000_{-0.0004}^{+0.0004}}
\newcommand{\BChiEffPost}{-0.0002_{-0.0004}^{+0.0003}}
\newcommand{\BMergerTimePost}{0.1_{-0.3}^{+0.3}}
\newcommand{\BSkyAreaPost}{0.0049}

\newcommand{\CPrimaryMassPost}{1.3506_{-0.0042}^{+0.0042}}
\newcommand{\CSecondaryMassPost}{0.4499_{-0.0012}^{+0.0012}}
\newcommand{\CRedshiftPost}{3.20_{-0.01}^{+0.01}}
\newcommand{\CChiEffPost}{-0.001_{-0.003}^{+0.003}}
\newcommand{\CMergerTimePost}{1_{-3}^{+4}}
\newcommand{\CSkyAreaPost}{25.7}

\newcommand{\DPrimaryMassPost}{53_{-3}^{+4}}
\newcommand{\DSecondaryMassPost}{17_{-1}^{+2}}
\newcommand{\DRedshiftPost}{2.0_{-0.2}^{+0.2}}
\newcommand{\DChiEffPost}{0.0_{-0.1}^{+0.1}}
\newcommand{\DMergerTimePost}{174_{-848}^{+658}}
\newcommand{\DSkyAreaPost}{13784}

\newcommand{\EPrimaryMassPost}{0.900_{-0.002}^{+0.001}}
\newcommand{\ESecondaryMassPost}{0.3000_{-0.0003}^{+0.0004}}
\newcommand{\ERedshiftPost}{2.299_{-0.003}^{+0.003}}
\newcommand{\EChiEffPost}{-0.001_{-0.001}^{+0.001}}
\newcommand{\EMergerTimePost}{0_{-1}^{+1}}
\newcommand{\ESkyAreaPost}{0.04}







\begin{table*}
\caption{\label{tab:PESummary}
Summary of results of our Bayesian parameter estimation for five binaries. The first column provides an ID for each binary, and the second, third, and fourth columns show the source frame component masses $m_1^{\rm inj}$, $m_2^{\rm inj}$ and redshift $z^{\rm inj}$, respectively. 
Consistent with earlier sections, we assume a mass ratio $q=1/3$.
All other parameters are identical for the five binaries: dimensionless spin magnitudes $\chi_1 = \chi_2 = \DimensionlessSpinOne$, ecliptic longitude $l=\EclipticLong$, 
sin of ecliptic latitude $\sin b = \sinEclipticLat$, 
inclination angle $\cos\iota = \cosInclination$, 
polarisation $\psi = \Polarization$, 
initial orbital phase $\phi=\InitialOrbitalPhase$, and merger time at $t_{\rm c} = \MergerTime\,{\rm s}$ from the start of the data.
The fifth and sixth columns summarise the SNR for these binaries with different multipole modes. The final six columns are the recovered posteriors for each binary, where $m_1$, $m_2$ are the source-frame component masses, $z$ is the redshift, $\chi_{\rm eff}$ is the aligned effective spin parameter, $\Omega_{90}$ is the $90^{\rm th}$ percentile of the (elliptical) sky area, and $\Delta t_{\rm c}$ is the recovered time of merger centred at the injected value. 
The values quoted with uncertainties are computed with all multipole modes, i.e., $\ell \geq 2$, and represent the median and central $90\%$ credible region for each parameter. Note that binaries 3 and 4 have multi-modal sky locations. 
Equal numbers of posterior samples from the \texttt{nessai} and \texttt{dynesty} analyses are used.
}
\vspace{0.1cm}
\setlength{\tabcolsep}{5pt}
\def\arraystretch{1.5}
\begin{tabular}{cccccc|cccccc}
\hline\hline
ID                 &
$m_1^{\rm inj}$    & 
$m_2^{\rm inj}$    &
$z^{\rm inj}$      & 
\multicolumn{2}{c|}{SNR ($f_{\rm low} = $ 0.1\,mHz)}  & 
$m_1$    & 
$m_2$    &
$z$                & 
$\chi_{\rm eff}$   &
$\Omega_{90}$     &
$\Delta t_{\rm c}$ \\

                 & 
$[10^6\,\Msol]$        & 
$[10^6\,\Msol]$        &
                 &       
$\ell = 2$ & $\ell \geq 2$ &
$[10^6\,\Msol]$        &
$[10^6\,\Msol]$        &
                 &
                 &
$[{\rm \deg}^2]$ &
$[{\rm s}]$      \\

\hline
1   &
$\BPrimaryMassInj$   & 
$\BSecondaryMassInj$ & 
$\BRedshiftInj$      & 
$\BSNRa$ & $\BSNRb$  &
$\BPrimaryMassPost$ & 
$\BSecondaryMassPost$ & 
$\BRedshiftPost$ & 
$\BChiEffPost$ &
$\BSkyAreaPost$ &
$\BMergerTimePost$        
\\

2   &
$\EPrimaryMassInj$   & 
$\ESecondaryMassInj$ & 
$\ERedshiftInj$      & 
$\ESNRa$ & $\ESNRb$  &
$\EPrimaryMassPost$ & 
$\ESecondaryMassPost$ & 
$\ERedshiftPost$ & 
$\EChiEffPost$ &
$\ESkyAreaPost$ &
$\EMergerTimePost$ 
\\

3   &
$\CPrimaryMassInj$   & 
$\CSecondaryMassInj$ & 
$\CRedshiftInj$      & 
$\CSNRa$ & $\CSNRb$  &
$\CPrimaryMassPost$ & 
$\CSecondaryMassPost$ & 
$\CRedshiftPost$ & 
$\CChiEffPost$ &
$\CSkyAreaPost$ &
$\CMergerTimePost$ 
\\

4   &
$\APrimaryMassInj$   & 
$\ASecondaryMassInj$ & 
$\ARedshiftInj$      & 
$\ASNRa$ & $\ASNRb$  & 
$\APrimaryMassPost$  &
$\ASecondaryMassPost$ & 
$\ARedshiftPost$ & 
$\AChiEffPost$ &
$\ASkyAreaPost$ &
$\AMergerTimePost$ 
\\

5   &
$\DPrimaryMassInj$   & 
$\DSecondaryMassInj$ & 
$\DRedshiftInj$      & 
$\DSNRa$ & $\DSNRb$  &
$\DPrimaryMassPost$ & 
$\DSecondaryMassPost$ & 
$\DRedshiftPost$ & 
$\DChiEffPost$ &
$\DSkyAreaPost$ &
$\DMergerTimePost$ 
\\

\hline\hline
\end{tabular}
\end{table*}

Lastly, we perform a Bayesian parameter estimation study of five representative MBHBs with the LISA inference tool \textsc{balrog}~\citep[see e.g.,][]{FinchEtAl:2022, KleinEtAl:2022, PrattenEtAl:2022, Buscicchio:2021, RoebberEtAl:2020} 
to provide an example of the quality of LISA observations for these systems. 

The five binaries are drawn randomly from the posterior distributions of the agnostic model in Section~\ref{sec:models} which provides the (source-frame) total mass $M$ and redshift $z$ of each binary. As the merger-rate posterior for the agnostic model highly disfavours high-mass binaries, but we are still interested in exploring the quality of LISA observations for such systems, we force the highest-mass draw to be detectable with $M > 10^7\,\Msol$. The properties of these systems are summarized in Table~\ref{tab:PESummary}. 

We compute the LISA noise-orthogonal time-delay-interferometry observables as described in Section~2 of~\cite{PrattenEtAl:2022}. 
Consistent with our analyses above, the signals are injected and recovered using the \textsc{IMRPhenomXHM} approximant~\citep{PrattenEtAl:2020,GarciaQuirosEtAl:2020}, and we convert between redshift and luminosity distance with the same cosmology as in Fig.~\ref{fig:Waterfall}. 
The injected (source frame) component masses ($m_1^{\rm inj}$, $m_2^{\rm inj}$) and redshifts $z^{\rm inj}$ for each of the binaries are shown in the left side of Table~\ref{tab:PESummary} along with the corresponding SNR we compute by either assuming only the $\ell = 2$ modes or the $\ell = 2$, 3, and 4 modes. 
We inject zero spins ($\chi_1 = \chi_2 = \DimensionlessSpinOne$) and mass ratio $q = 1/3$ for all five binaries, implying that the $\chi_{\rm eff}$ priors are centered on 0. The remaining extrinsic and intrinsic parameters are identical for each binary and are summarised in the caption of Table~\ref{tab:PESummary}. 
We assume a LISA configuration with $2.5\,\text{million-km}$ arm length and a data duration of $4\,{\rm yr}$,  
and for recovery we use two implementations of nested sampling 
\citep{SkillingNestedSampling:2006}: 
\textsc{dynesty}~\citep{SpeagleDynesty:2020,dynesty} and \textsc{nessai}~\citep{WilliamsEtAlNessai:2021,nessai}.

\begin{figure*}
\begin{center}
\includegraphics[width=1\textwidth]{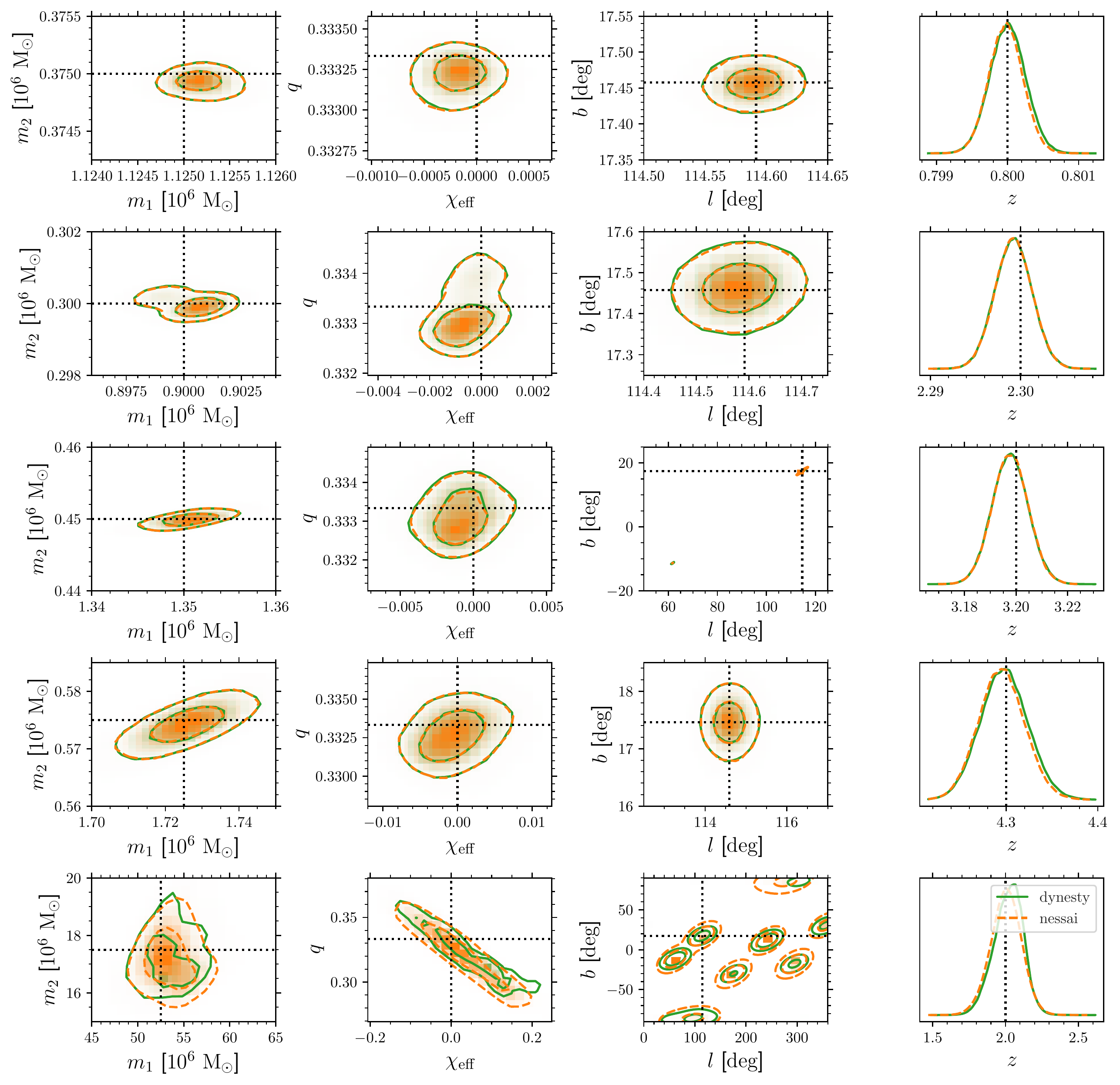}
\end{center}
\caption{\label{fig:PEresults}
    Density estimation of the posterior distributions for the five binaries in our Bayesian parameter estimation of Section~\ref{sec:PE}. The \textsc{dynesty} (\textsc{nessai}) results are indicated by the solid green (dashed orange) lines. Each row corresponds to one of the binaries, i.e., binaries 1-5 from top to bottom, and the four columns correspond to the recovered source-frame component masses $m_1$ and $m_2$, the mass ratio $q$ and aligned effective spin $\chi_{\rm eff}$, the ecliptic longitude $l$ and latitude $b$, and the redshift $z$, respectively. }
\end{figure*}

Fig.~\ref{fig:PEresults}
shows selected posterior distributions, where the five rows top-to-bottom correspond to the five binaries, and shows a comparison between the two samplers. 
These results are summarised in Table~\ref{tab:PESummary}, where the first column provides an ID for each binary, the next five columns specify the injected values and the corresponding SNRs using only the leading modes or all available modes, and the remaining six columns list the medians and central $90\%$ credible regions of the recovered posterior distributions. These quoted values are computed from equally mixed samples of the \textsc{dynesty} and \textsc{nessai} results shown in Fig.~\ref{fig:PEresults}. 

The five binaries are listed in Table~\ref{tab:PESummary} by decreasing SNRs, which span a broad range. 
Intuitively, Binary 1 with the highest SNR displays the smallest recovered parameter uncertainties, and is closely followed by Binaries 2 and 3. Binary 4, which was injected with the furthest distance, has an order-of-magnitude smaller SNR and corresponding parameters that are recovered with larger uncertainties. Nevertheless, its masses and spins are still precisely measured. While we are confident in the sky locations of these binaries, i.e. the panels in the third column of Fig.~\ref{fig:PEresults}, these skymaps are complicated as they can suffer from degeneracies and as more time spent in the LISA band can suppress secondary skymodes \citep{PrattenEtAl:2022}, and investigating them further is beyond the scope of this work. 
Still, the sky areas of these four binaries will be sufficiently small to support realistic electromagnetic follow-up campaigns \citep{MangiagliEtAl:2022}. 

Importantly, Binary 5 stands out among the others as it has the largest injected total mass, and hence the lowest SNR. In the context of the three mass-redshift bins of Table~\ref{tab:DRSummary}, this is the only binary of the five that lies in a higher mass region (mid bin), implying its SNR may be sensitive to the modelling and instrumental assumptions we explored earlier. Indeed, we find that its SNR is nearly insufficient, i.e., $\approx 8$, to be detectable unless we include higher modes, as shown in the fifth and sixth columns of Table~\ref{tab:PESummary}, or assume a smaller LISA low-frequency limit than $10^{-4}$ Hz. Consequently, compared to the other binaries, its recovered parameters suffer from significantly larger uncertainties and its sky location is burdened by multi-modality, i.e., see the panel in the fifth row and third column of Fig.~\ref{fig:PEresults}.

These results demonstrate LISA's exceptional potential to measure the properties of MBHBs.

\section{Conclusions \& Discussion}
\label{sec:Disc}

The future GW detector LISA will observe MBHBs with remarkable precision due to the high SNR it will achieve across large ranges of redshift and mass. In this work, we have demonstrated how PTA measurements of the SGWB can inform the potential of detecting MBHBs with LISA. To do this we constrained MBHB formation models to obtain estimates of the merger rates of the MBHB population, and then we computed the SNR of these MBHBs to arrive at their LISA detection rates. We also performed a parameter estimation study of a handful of such binaries to showcase the tremendous constraining power of LISA. Our findings are summarized in these key conclusions:
\begin{enumerate}
    \item Despite primarily probing different portions of the parameter space of MBHBs, PTAs and LISA can jointly provide robust predictions for MBHBs.

    \item The astrophysical assumptions of the two formation models we consider can lead to different predictions for the merger rate, and hence detection rate, of MBHBs.

    \item Our LISA detection rates $\mathcal{R}$ for binaries with mass $M \gtrsim 10^6\,\Msol$ decrease monotonically with increasing M, which parallels the SNR of LISA, and are boosted by higher modes and a small LISA low-frequency limit. 
     
    \item Binaries with higher mass, i.e., $M \gtrsim 10^7\,\Msol$, and near the edge of the LISA horizon can be undetectable without these boosting effects, e.g., $\mathcal{R} \approx 0(0.1) - 6(11)\,\mathrm{yr}^{-1}$ for agnostic (astro-informed) models at central $90$\% credible interval, but $\mathcal{R}$ can quickly vanish with a pessimistic LISA low-frequency limit.

    \item A long mission duration for LISA helps to ensure detection of high-mass binaries when $\mathcal{R} \sim 0.1\,\mathrm{yr}^{-1}$.

    \item LISA's ability to adequately measure the parameters of high-mass binaries will rely heavily on modelling assumptions (such as including higher modes) and instrumental assumptions (such as the low-frequency limit).
\end{enumerate}
Although the MBHB formation models we consider are uncertain and despite the current challenges with PTA measurements of the SGWB, constraints on MBHB merger rates with PTAs can be used to make meaningful predictions for LISA observations. 
The present work is a proof-of-principle that multi-band studies of MBHBs are advantageous and offer a viable probe of the MBHB population. 

Precise sky location estimates from GWs are of particular importance for multi-messenger, i.e., joint GW and electromagnetic, observations of MBHBs, see e.g.~\citet{Piro:2023}. 
As higher modes are known to break degeneracies 
and provide multi-modal sky localisations \citep{Marsat:2021,PrattenEtAl:2022}, 
including higher modes will be important for multi-messenger detections, e.g., of bright quasars with high mass $M \gtrsim 10^8\,\Msol$ and redshift $z \gtrsim 6$ \citep{Volonteri:2021}.

There are a few caveats in our analysis worth discussion. 
We want to emphasise that, to illustrate the point concerning the synergy 
between PTAs and LISA, we used an ansatz of assuming the common red-stochastic 
signal observed in PTA data is due to a SGWB from MBHBs. 
The nature of this signal is currently unknown, and there is no statistically significant evidence that it is due to a SGWB, e.g.'s see~\citet{NANOGrav12p5yr:2020, PPTA:2021, EPTA:2021, IPTA:2022} for detailed discussions. Moreover, if a SGWB is detected by PTAs in the future the physical source of the signal would need to be identified, such as an astrophysical population of MBHBs or some other (possibly more exotic) process in the early Universe, see e.g.'s~\cite{2021NatAs...5.1268M, 2021PhRvL.127y1302A, 2021PhRvL.127y1303X}. These are highly non-trivial problems that we do not consider here. 
However, under the assumption of a ``universal'' MBHB population in the Universe, the results of the LISA survey may provide the strongest clue in this direction and further the relationship between PTAs and LISA.

Throughout this work we have assumed that the MBHBs are non-spinning. Although we do not explore it here, we expect that our MBHBs detection rates will be most sensitive to the spin magnitudes and directions in the high-mass regime, i.e., $M \gtrsim 10^7\,\Msol$, as their LISA SNR is more sensitive to higher modes than lower mass binaries. Lastly, the detection rates that we compute are limited by the uncertainties of the two formation models that we consider. Probing astrophysical observables is challenging with present PTA datasets \citep{ChenSesanaConselice:2019}, and further work is needed in the modelling of MBHB populations. 

Frameworks such as ours that attempt to forecast the science potential of LISA using SGWB measurements from PTAs should improve in the near future as PTAs continue the search for this signal. Indeed, several PTA consortia have very recently announced emerging evidence for a GW signal at the 2$\sigma$ to 4$\sigma$ level \citep{ETPAGWBAntoniadisEtAl:2023,PPTAGWReardonEtAl:2023,NANOGravGWBAgazieEtAl:2023,CPTAGWBXuEtAl:2023}, and although the source of the signal remains uncertain, the growing evidence for detection adds support to a primary assumption of our framework and invites more detailed studies. Equally, once operational, LISA's observations of individual MBHBs at lower masses will aid PTAs in constraining the MBHB population at high masses, even if difficulties in detecting a SGWB from MBHBs persist. 
These complementary observations, as well as those at high LISA frequencies~\citep{KleinEtAl:2022}, will help enable a multi-band era of GW astrophysics. 

\section*{Acknowledgements}
The Authors thank Diganta Bandopadhyay for their contributions in the interpretation of our results, and the anonymous referee for insightful comments.
N.S. is supported by the Leverhulme Trust Grant No. RPG-2019-350. %
A.K., H.M., C.J.M., and A.V. acknowledge the support of the UK Space Agency, Grant No. ST/V002813/1 and ST/X002071/1. GP gratefully acknowledges support from a Royal Society University Research Fellowship URF{\textbackslash}R1{\textbackslash}221500 and RF{\textbackslash}ERE{\textbackslash}221015, and STFC grant ST/V005677/1.
R.B. acknowledges support through the Italian Space Agency grant \emph{Phase A activity for LISA mission, Agreement n. 2017-29-H.0, CUP F62F17000290005.
}
A.V. acknowledges the support of the Royal Society and Wolfson Foundation. 
Some of the computations described in this paper were performed using the University of Birmingham's BlueBEAR HPC service, which provides a High Performance Computing service to the University's research community. See~\url{http://www.birmingham.ac.uk/bear} for more details.
Parts of this work were performed on the OzSTAR national facility at Swinburne University of Technology. 
The OzSTAR program receives funding in part from the Astronomy National Collaborative Research Infrastructure Strategy (NCRIS) allocation provided by the Australian Government.
The Authors thank all the developers of the MBHB models and of the \textsc{balrog} codesuite, including those who are not authors here.
Besides the software tools cited in the main text, this work has made use of \textsc{astropy}~\citep{astropy:2022}, \textsc{corner}~\citep{corner}, \textsc{matplotlib}~\citep{matplotlib}, \textsc{numpy}~\citep{numpy}, \textsc{scipy}~\citep{scipy}, and \textsc{seaborn}~\citep{seaborn}.

\section*{Data Availability}
The data underlying this article will be shared on reasonable request to the correspondence author.

\bibliographystyle{mnras}
\bibliography{refs}{}

\begin{thebibliography}{}
\makeatletter
\relax
\def\mn@urlcharsother{\let\do\@makeother \do\$\do\&\do\#\do\^\do\_\do\%\do\~}
\def\mn@doi{\begingroup\mn@urlcharsother \@ifnextchar [ {\mn@doi@}
  {\mn@doi@[]}}
\def\mn@doi@[#1]#2{\def\@tempa{#1}\ifx\@tempa\@empty \href
  {http://dx.doi.org/#2} {doi:#2}\else \href {http://dx.doi.org/#2} {#1}\fi
  \endgroup}
\def\mn@eprint#1#2{\mn@eprint@#1:#2::\@nil}
\def\mn@eprint@arXiv#1{\href {http://arxiv.org/abs/#1} {{\tt arXiv:#1}}}
\def\mn@eprint@dblp#1{\href {http://dblp.uni-trier.de/rec/bibtex/#1.xml}
  {dblp:#1}}
\def\mn@eprint@#1:#2:#3:#4\@nil{\def\@tempa {#1}\def\@tempb {#2}\def\@tempc
  {#3}\ifx \@tempc \@empty \let \@tempc \@tempb \let \@tempb \@tempa \fi \ifx
  \@tempb \@empty \def\@tempb {arXiv}\fi \@ifundefined
  {mn@eprint@\@tempb}{\@tempb:\@tempc}{\expandafter \expandafter \csname
  mn@eprint@\@tempb\endcsname \expandafter{\@tempc}}}

\bibitem[\protect\citeauthoryear{{Afzal} et~al.,}{{Afzal}
  et~al.}{2023}]{NANOGravImplicationsNewPhysics:2023}
{Afzal} A.,  et~al., 2023, \mn@doi [\apjl] {10.3847/2041-8213/acdc91}, \href
  {https://ui.adsabs.harvard.edu/abs/2023ApJ...951L..11A} {951, L11}

\bibitem[\protect\citeauthoryear{{Agazie} et~al.,}{{Agazie}
  et~al.}{2023a}]{NANOGravImplicationsMBHB:2023}
{Agazie} G.,  et~al., 2023a, \mn@doi [arXiv e-prints]
  {10.48550/arXiv.2306.16220}, \href
  {https://ui.adsabs.harvard.edu/abs/2023arXiv230616220A} {p. arXiv:2306.16220}

\bibitem[\protect\citeauthoryear{{Agazie} et~al.,}{{Agazie}
  et~al.}{2023b}]{NANOGravGWBAgazieEtAl:2023}
{Agazie} G.,  et~al., 2023b, \mn@doi [\apjl] {10.3847/2041-8213/acdac6}, \href
  {https://ui.adsabs.harvard.edu/abs/2023ApJ...951L...8A} {951, L8}

\bibitem[\protect\citeauthoryear{{Amaro-Seoane} et~al.,}{{Amaro-Seoane}
  et~al.}{2017}]{Amaro-Seoane:2017}
{Amaro-Seoane} P.,  et~al., 2017, arXiv e-prints, \href
  {https://ui.adsabs.harvard.edu/abs/2017arXiv170200786A} {p. arXiv:1702.00786}

\bibitem[\protect\citeauthoryear{{Amaro-Seoane}, {Andrews}, {Arca Sedda}
  et~al.}{{Amaro-Seoane} et~al.}{2023}]{Amaro-Seoane:2023}
{Amaro-Seoane} P.,  {Andrews} J.,  {Arca Sedda} M.,   et~al., 2023, \mn@doi
  [Living Reviews in Relativity] {10.1007/s41114-022-00041-y}, \href
  {https://ui.adsabs.harvard.edu/abs/2023LRR....26....2A} {26, 2}

\bibitem[\protect\citeauthoryear{{Antoniadis} et~al.,}{{Antoniadis}
  et~al.}{2022}]{IPTA:2022}
{Antoniadis} J.,  et~al., 2022, \mn@doi [\mnras] {10.1093/mnras/stab3418},
  \href {https://ui.adsabs.harvard.edu/abs/2022MNRAS.510.4873A} {510, 4873}

\bibitem[\protect\citeauthoryear{{Antoniadis} et~al.,}{{Antoniadis}
  et~al.}{2023a}]{ETPAGWBAntoniadisEtAl:2023}
{Antoniadis} J.,  et~al., 2023a, \mn@doi [arXiv e-prints]
  {10.48550/arXiv.2306.16214}, \href
  {https://ui.adsabs.harvard.edu/abs/2023arXiv230616214A} {p. arXiv:2306.16214}

\bibitem[\protect\citeauthoryear{{Antoniadis} et~al.,}{{Antoniadis}
  et~al.}{2023b}]{EPTAImplications:2023}
{Antoniadis} J.,  et~al., 2023b, \mn@doi [arXiv e-prints]
  {10.48550/arXiv.2306.16227}, \href
  {https://ui.adsabs.harvard.edu/abs/2023arXiv230616227A} {p. arXiv:2306.16227}

\bibitem[\protect\citeauthoryear{{Arzoumanian} et~al.,}{{Arzoumanian}
  et~al.}{2020}]{NANOGrav12p5yr:2020}
{Arzoumanian} Z.,  et~al., 2020, \mn@doi [\apjl] {10.3847/2041-8213/abd401},
  \href {https://ui.adsabs.harvard.edu/abs/2020ApJ...905L..34A} {905, L34}

\bibitem[\protect\citeauthoryear{{Arzoumanian} et~al.,}{{Arzoumanian}
  et~al.}{2021}]{2021PhRvL.127y1302A}
{Arzoumanian} Z.,  et~al., 2021, \mn@doi [\prl]
  {10.1103/PhysRevLett.127.251302}, \href
  {https://ui.adsabs.harvard.edu/abs/2021PhRvL.127y1302A} {127, 251302}

\bibitem[\protect\citeauthoryear{{Astropy Collaboration} et~al.,}{{Astropy
  Collaboration} et~al.}{2022}]{astropy:2022}
{Astropy Collaboration} et~al., 2022, \mn@doi [apj] {10.3847/1538-4357/ac7c74},
  \href {https://ui.adsabs.harvard.edu/abs/2022ApJ...935..167A} {935, 167}

\bibitem[\protect\citeauthoryear{{Auclair} et~al.,}{{Auclair}
  et~al.}{2022}]{AuclairEtAl:2022}
{Auclair} P.,  et~al., 2022, \mn@doi [arXiv e-prints]
  {10.48550/arXiv.2204.05434}, \href
  {https://ui.adsabs.harvard.edu/abs/2022arXiv220405434A} {p. arXiv:2204.05434}

\bibitem[\protect\citeauthoryear{{Babak} et~al.,}{{Babak}
  et~al.}{2017}]{BabakEtAl:2017}
{Babak} S.,  et~al., 2017, \mn@doi [\prd]
  {10.1103/PhysRevD.95.10301210.48550/arXiv.1703.09722}, \href
  {https://ui.adsabs.harvard.edu/abs/2017PhRvD..95j3012B} {95, 103012}

\bibitem[\protect\citeauthoryear{{Babak}, {Hewitson}  \& {Petiteau}}{{Babak}
  et~al.}{2021}]{BabakEtAl:2021}
{Babak} S.,  {Hewitson} M.,   {Petiteau} A.,  2021, \mn@doi [arXiv e-prints]
  {10.48550/arXiv.2108.01167}, \href
  {https://ui.adsabs.harvard.edu/abs/2021arXiv210801167B} {p. arXiv:2108.01167}

\bibitem[\protect\citeauthoryear{{Bailes} et~al.,}{{Bailes}
  et~al.}{2021}]{BailesEtAl:2021}
{Bailes} M.,  et~al., 2021, \mn@doi [Nature Reviews Physics]
  {10.1038/s42254-021-00303-8}, \href
  {https://ui.adsabs.harvard.edu/abs/2021NatRP...3..344B} {3, 344}

\bibitem[\protect\citeauthoryear{{Barack} \& {Cutler}}{{Barack} \&
  {Cutler}}{2004}]{Barack:2004}
{Barack} L.,  {Cutler} C.,  2004, \mn@doi [\prd] {10.1103/PhysRevD.69.082005},
  \href {https://ui.adsabs.harvard.edu/abs/2004PhRvD..69h2005B} {69, 082005}

\bibitem[\protect\citeauthoryear{{Barausse}, {Dvorkin}, {Tremmel}, {Volonteri}
  \& {Bonetti}}{{Barausse} et~al.}{2020}]{Barausse:2020}
{Barausse} E.,  {Dvorkin} I.,  {Tremmel} M.,  {Volonteri} M.,   {Bonetti} M.,
  2020, \mn@doi [\apj] {10.3847/1538-4357/abba7f}, \href
  {https://ui.adsabs.harvard.edu/abs/2020ApJ...904...16B} {904, 16}

\bibitem[\protect\citeauthoryear{{Bonetti}, {Sesana}, {Haardt}, {Barausse}  \&
  {Colpi}}{{Bonetti} et~al.}{2019}]{Bonetti:2019}
{Bonetti} M.,  {Sesana} A.,  {Haardt} F.,  {Barausse} E.,   {Colpi} M.,  2019,
  \mn@doi [\mnras] {10.1093/mnras/stz903}, \href
  {https://ui.adsabs.harvard.edu/abs/2019MNRAS.486.4044B} {486, 4044}

\bibitem[\protect\citeauthoryear{{Buscicchio}, {Klein}, {Roebber}, {Moore},
  {Gerosa}, {Finch}  \& {Vecchio}}{{Buscicchio} et~al.}{2021}]{Buscicchio:2021}
{Buscicchio} R.,  {Klein} A.,  {Roebber} E.,  {Moore} C.~J.,  {Gerosa} D.,
  {Finch} E.,   {Vecchio} A.,  2021, \mn@doi [\prd]
  {10.1103/PhysRevD.104.044065}, \href
  {https://ui.adsabs.harvard.edu/abs/2021PhRvD.104d4065B} {104, 044065}

\bibitem[\protect\citeauthoryear{{Chen}, {Middleton}, {Sesana}, {Del Pozzo}  \&
  {Vecchio}}{{Chen} et~al.}{2017a}]{ChenEtAl:2017}
{Chen} S.,  {Middleton} H.,  {Sesana} A.,  {Del Pozzo} W.,   {Vecchio} A.,
  2017a, \mn@doi [\mnras] {10.1093/mnras/stx475}, \href
  {https://ui.adsabs.harvard.edu/abs/2017MNRAS.468..404C} {468, 404}

\bibitem[\protect\citeauthoryear{{Chen}, {Sesana}  \& {Del Pozzo}}{{Chen}
  et~al.}{2017b}]{ChenSesanaDelPozzo:2017}
{Chen} S.,  {Sesana} A.,   {Del Pozzo} W.,  2017b, \mn@doi [\mnras]
  {10.1093/mnras/stx1093}, \href
  {https://ui.adsabs.harvard.edu/abs/2017MNRAS.470.1738C} {470, 1738}

\bibitem[\protect\citeauthoryear{{Chen}, {Sesana}  \& {Conselice}}{{Chen}
  et~al.}{2019}]{ChenSesanaConselice:2019}
{Chen} S.,  {Sesana} A.,   {Conselice} C.~J.,  2019, \mn@doi [\mnras]
  {10.1093/mnras/stz1722}, \href
  {https://ui.adsabs.harvard.edu/abs/2019MNRAS.488..401C} {488, 401}

\bibitem[\protect\citeauthoryear{{Chen}, {Yu}  \& {Lu}}{{Chen}
  et~al.}{2020}]{ChenYuAndLu:2020}
{Chen} Y.,  {Yu} Q.,   {Lu} Y.,  2020, \mn@doi [\apj]
  {10.3847/1538-4357/ab9594}, \href
  {https://ui.adsabs.harvard.edu/abs/2020ApJ...897...86C} {897, 86}

\bibitem[\protect\citeauthoryear{{Chen} et~al.,}{{Chen}
  et~al.}{2021}]{EPTA:2021}
{Chen} S.,  et~al., 2021, \mn@doi [\mnras] {10.1093/mnras/stab2833}, \href
  {https://ui.adsabs.harvard.edu/abs/2021MNRAS.508.4970C} {508, 4970}

\bibitem[\protect\citeauthoryear{{Cutler} \& {Flanagan}}{{Cutler} \&
  {Flanagan}}{1994}]{Cutler:1994}
{Cutler} C.,  {Flanagan} {\'E}.~E.,  1994, \mn@doi [\prd]
  {10.1103/PhysRevD.49.2658}, \href
  {https://ui.adsabs.harvard.edu/abs/1994PhRvD..49.2658C} {49, 2658}

\bibitem[\protect\citeauthoryear{Ellis \& van Haasteren}{Ellis \& van
  Haasteren}{2017}]{ptmcmc}
Ellis J.,  van Haasteren R.,  2017, jellis18/PTMCMCSampler: Official Release,
  \mn@doi{10.5281/zenodo.1037578}, \url
  {https://doi.org/10.5281/zenodo.1037578}

\bibitem[\protect\citeauthoryear{{Ellis}, {Fairbairn}, {H{\"u}tsi}, {Raidal},
  {Urrutia}, {Vaskonen}  \& {Veerm{\"a}e}}{{Ellis}
  et~al.}{2023}]{EllisEtAl:2023}
{Ellis} J.,  {Fairbairn} M.,  {H{\"u}tsi} G.,  {Raidal} M.,  {Urrutia} J.,
  {Vaskonen} V.,   {Veerm{\"a}e} H.,  2023, \mn@doi [arXiv e-prints]
  {10.48550/arXiv.2301.13854}, \href
  {https://ui.adsabs.harvard.edu/abs/2023arXiv230113854E} {p. arXiv:2301.13854}

\bibitem[\protect\citeauthoryear{{Finch} et~al.,}{{Finch}
  et~al.}{2022}]{FinchEtAl:2022}
{Finch} E.,  et~al., 2022, \mn@doi [arXiv e-prints]
  {10.48550/arXiv.2210.10812}, \href
  {https://ui.adsabs.harvard.edu/abs/2022arXiv221010812F} {p. arXiv:2210.10812}

\bibitem[\protect\citeauthoryear{Foreman-Mackey}{Foreman-Mackey}{2016}]{corner}
Foreman-Mackey D.,  2016, \mn@doi [The Journal of Open Source Software]
  {10.21105/joss.00024}, 1, 24

\bibitem[\protect\citeauthoryear{{Foster} \& {Backer}}{{Foster} \&
  {Backer}}{1990}]{FosterBacker:1990}
{Foster} R.~S.,  {Backer} D.~C.,  1990, \mn@doi [\apj] {10.1086/169195}, \href
  {https://ui.adsabs.harvard.edu/abs/1990ApJ...361..300F} {361, 300}

\bibitem[\protect\citeauthoryear{{Garc{\'\i}a-Quir{\'o}s}, {Colleoni}, {Husa},
  {Estell{\'e}s}, {Pratten}, {Ramos-Buades}, {Mateu-Lucena}  \&
  {Jaume}}{{Garc{\'\i}a-Quir{\'o}s} et~al.}{2020}]{GarciaQuirosEtAl:2020}
{Garc{\'\i}a-Quir{\'o}s} C.,  {Colleoni} M.,  {Husa} S.,  {Estell{\'e}s} H.,
  {Pratten} G.,  {Ramos-Buades} A.,  {Mateu-Lucena} M.,   {Jaume} R.,  2020,
  \mn@doi [\prd] {10.1103/PhysRevD.102.064002}, \href
  {https://ui.adsabs.harvard.edu/abs/2020PhRvD.102f4002G} {102, 064002}

\bibitem[\protect\citeauthoryear{{Goncharov} et~al.,}{{Goncharov}
  et~al.}{2021}]{PPTA:2021}
{Goncharov} B.,  et~al., 2021, \mn@doi [\apjl] {10.3847/2041-8213/ac17f4},
  \href {https://ui.adsabs.harvard.edu/abs/2021ApJ...917L..19G} {917, L19}

\bibitem[\protect\citeauthoryear{Harris et~al.,}{Harris et~al.}{2020}]{numpy}
Harris C.~R.,  et~al., 2020, \mn@doi [Nature] {10.1038/s41586-020-2649-2}, 585,
  357

\bibitem[\protect\citeauthoryear{{Heckman} \& {Best}}{{Heckman} \&
  {Best}}{2014}]{Heckman:2014}
{Heckman} T.~M.,  {Best} P.~N.,  2014, \mn@doi [\araa]
  {10.1146/annurev-astro-081913-035722}, \href
  {https://ui.adsabs.harvard.edu/abs/2014ARA&A..52..589H} {52, 589}

\bibitem[\protect\citeauthoryear{{Hogg}}{{Hogg}}{1999}]{Hogg:1999}
{Hogg} D.~W.,  1999, arXiv e-prints, \href
  {https://ui.adsabs.harvard.edu/abs/1999astro.ph..5116H} {pp
  astro--ph/9905116}

\bibitem[\protect\citeauthoryear{Hunter}{Hunter}{2007}]{matplotlib}
Hunter J.~D.,  2007, \mn@doi [Computing in Science \& Engineering]
  {10.1109/MCSE.2007.55}, 9, 90

\bibitem[\protect\citeauthoryear{{Katz} \& {Larson}}{{Katz} \&
  {Larson}}{2019}]{Katz:2019}
{Katz} M.~L.,  {Larson} S.~L.,  2019, \mn@doi [\mnras] {10.1093/mnras/sty3321},
  \href {https://ui.adsabs.harvard.edu/abs/2019MNRAS.483.3108K} {483, 3108}

\bibitem[\protect\citeauthoryear{{Katz}, {Kelley}, {Dosopoulou}, {Berry},
  {Blecha}  \& {Larson}}{{Katz} et~al.}{2020}]{Katz2020}
{Katz} M.~L.,  {Kelley} L.~Z.,  {Dosopoulou} F.,  {Berry} S.,  {Blecha} L.,
  {Larson} S.~L.,  2020, \mn@doi [\mnras] {10.1093/mnras/stz3102}, \href
  {https://ui.adsabs.harvard.edu/abs/2020MNRAS.491.2301K} {491, 2301}

\bibitem[\protect\citeauthoryear{{Kelley}, {Blecha}, {Hernquist}, {Sesana}  \&
  {Taylor}}{{Kelley} et~al.}{2017}]{KelleyEtAl:2017b}
{Kelley} L.~Z.,  {Blecha} L.,  {Hernquist} L.,  {Sesana} A.,   {Taylor} S.~R.,
  2017, \mn@doi [\mnras] {10.1093/mnras/stx1638}, \href
  {https://ui.adsabs.harvard.edu/abs/2017MNRAS.471.4508K} {471, 4508}

\bibitem[\protect\citeauthoryear{{Klein} et~al.,}{{Klein}
  et~al.}{2016}]{Klein:2016}
{Klein} A.,  et~al., 2016, \mn@doi [\prd] {10.1103/PhysRevD.93.024003}, \href
  {https://ui.adsabs.harvard.edu/abs/2016PhRvD..93b4003K} {93, 024003}

\bibitem[\protect\citeauthoryear{{Klein} et~al.,}{{Klein}
  et~al.}{2022}]{KleinEtAl:2022}
{Klein} A.,  et~al., 2022, \mn@doi [arXiv e-prints]
  {10.48550/arXiv.2204.03423}, \href
  {https://ui.adsabs.harvard.edu/abs/2022arXiv220403423K} {p. arXiv:2204.03423}

\bibitem[\protect\citeauthoryear{Koposov et~al.,}{Koposov
  et~al.}{2022}]{dynesty}
Koposov S.,  et~al., 2022, joshspeagle/dynesty: v2.0.3,
  \mn@doi{10.5281/zenodo.7388523}, \url
  {https://doi.org/10.5281/zenodo.7388523}

\bibitem[\protect\citeauthoryear{{Kormendy} \& {Ho}}{{Kormendy} \&
  {Ho}}{2013}]{Kormendy:2013}
{Kormendy} J.,  {Ho} L.~C.,  2013, \mn@doi [\araa]
  {10.1146/annurev-astro-082708-101811}, \href
  {https://ui.adsabs.harvard.edu/abs/2013ARA&A..51..511K} {51, 511}

\bibitem[\protect\citeauthoryear{{Kormendy} \& {Richstone}}{{Kormendy} \&
  {Richstone}}{1995}]{Kormendy:1995}
{Kormendy} J.,  {Richstone} D.,  1995, \mn@doi [\araa]
  {10.1146/annurev.aa.33.090195.003053}, \href
  {https://ui.adsabs.harvard.edu/abs/1995ARA&A..33..581K} {33, 581}

\bibitem[\protect\citeauthoryear{{Mangiagli}, {Caprini}, {Volonteri}, {Marsat},
  {Vergani}, {Tamanini}  \& {Inchausp{\'e}}}{{Mangiagli}
  et~al.}{2022}]{MangiagliEtAl:2022}
{Mangiagli} A.,  {Caprini} C.,  {Volonteri} M.,  {Marsat} S.,  {Vergani} S.,
  {Tamanini} N.,   {Inchausp{\'e}} H.,  2022, \mn@doi [\prd]
  {10.1103/PhysRevD.106.103017}, \href
  {https://ui.adsabs.harvard.edu/abs/2022PhRvD.106j3017M} {106, 103017}

\bibitem[\protect\citeauthoryear{{Marsat}, {Baker}  \& {Canton}}{{Marsat}
  et~al.}{2021}]{Marsat:2021}
{Marsat} S.,  {Baker} J.~G.,   {Canton} T.~D.,  2021, \mn@doi [\prd]
  {10.1103/PhysRevD.103.083011}, \href
  {https://ui.adsabs.harvard.edu/abs/2021PhRvD.103h3011M} {103, 083011}

\bibitem[\protect\citeauthoryear{{Middleton}, {Del Pozzo}, {Farr}, {Sesana}  \&
  {Vecchio}}{{Middleton} et~al.}{2016}]{MiddletonEtAl:2016}
{Middleton} H.,  {Del Pozzo} W.,  {Farr} W.~M.,  {Sesana} A.,   {Vecchio} A.,
  2016, \mn@doi [\mnras] {10.1093/mnrasl/slv150}, \href
  {https://ui.adsabs.harvard.edu/abs/2016MNRAS.455L..72M} {455, L72}

\bibitem[\protect\citeauthoryear{{Middleton}, {Chen}, {Del Pozzo}, {Sesana}  \&
  {Vecchio}}{{Middleton} et~al.}{2018}]{MiddEtAl:2018}
{Middleton} H.,  {Chen} S.,  {Del Pozzo} W.,  {Sesana} A.,   {Vecchio} A.,
  2018, \mn@doi [Nature Communications] {10.1038/s41467-018-02916-7}, \href
  {https://ui.adsabs.harvard.edu/abs/2018NatCo...9..573M} {9, 573}

\bibitem[\protect\citeauthoryear{{Middleton}, {Sesana}, {Chen}, {Vecchio}, {Del
  Pozzo}  \& {Rosado}}{{Middleton} et~al.}{2021}]{MiddEtAl:2021}
{Middleton} H.,  {Sesana} A.,  {Chen} S.,  {Vecchio} A.,  {Del Pozzo} W.,
  {Rosado} P.~A.,  2021, \mn@doi [\mnras] {10.1093/mnrasl/slab008}, \href
  {https://ui.adsabs.harvard.edu/abs/2021MNRAS.502L..99M} {502, L99}

\bibitem[\protect\citeauthoryear{{Moore} \& {Vecchio}}{{Moore} \&
  {Vecchio}}{2021}]{2021NatAs...5.1268M}
{Moore} C.~J.,  {Vecchio} A.,  2021, \mn@doi [Nature Astronomy]
  {10.1038/s41550-021-01489-8}, \href
  {https://ui.adsabs.harvard.edu/abs/2021NatAs...5.1268M} {5, 1268}

\bibitem[\protect\citeauthoryear{{Moore}, {Cole}  \& {Berry}}{{Moore}
  et~al.}{2015}]{Moore:2015}
{Moore} C.~J.,  {Cole} R.~H.,   {Berry} C.~P.~L.,  2015, \mn@doi [Classical and
  Quantum Gravity] {10.1088/0264-9381/32/1/015014}, \href
  {https://ui.adsabs.harvard.edu/abs/2015CQGra..32a5014M} {32, 015014}

\bibitem[\protect\citeauthoryear{{Phinney}}{{Phinney}}{2001}]{Phinney:2001}
{Phinney} E.~S.,  2001, arXiv e-prints, \href
  {https://ui.adsabs.harvard.edu/abs/2001astro.ph..8028P} {pp
  astro--ph/0108028}

\bibitem[\protect\citeauthoryear{{Piro}, {Colpi}, {Aird}  et~al.}{{Piro}
  et~al.}{2023}]{Piro:2023}
{Piro} L.,  {Colpi} M.,  {Aird} J.,   et~al., 2023, \mn@doi [\mnras]
  {10.1093/mnras/stad659}, \href
  {https://ui.adsabs.harvard.edu/abs/2023MNRAS.521.2577P} {521, 2577}

\bibitem[\protect\citeauthoryear{{Planck Collaboration} et~al.,}{{Planck
  Collaboration} et~al.}{2020}]{Planck:2018}
{Planck Collaboration} et~al., 2020, \mn@doi [\aap]
  {10.1051/0004-6361/201833910}, \href
  {https://ui.adsabs.harvard.edu/abs/2020A&A...641A...6P} {641, A6}

\bibitem[\protect\citeauthoryear{{Pratten}, {Husa}, {Garc{\'\i}a-Quir{\'o}s},
  {Colleoni}, {Ramos-Buades}, {Estell{\'e}s}  \& {Jaume}}{{Pratten}
  et~al.}{2020}]{PrattenEtAl:2020}
{Pratten} G.,  {Husa} S.,  {Garc{\'\i}a-Quir{\'o}s} C.,  {Colleoni} M.,
  {Ramos-Buades} A.,  {Estell{\'e}s} H.,   {Jaume} R.,  2020, \mn@doi [\prd]
  {10.1103/PhysRevD.102.064001}, \href
  {https://ui.adsabs.harvard.edu/abs/2020PhRvD.102f4001P} {102, 064001}

\bibitem[\protect\citeauthoryear{{Pratten}, {Klein}, {Moore}, {Middleton},
  {Steinle}, {Schmidt}  \& {Vecchio}}{{Pratten}
  et~al.}{2022}]{PrattenEtAl:2022}
{Pratten} G.,  {Klein} A.,  {Moore} C.~J.,  {Middleton} H.,  {Steinle} N.,
  {Schmidt} P.,   {Vecchio} A.,  2022, arXiv e-prints, \href
  {https://ui.adsabs.harvard.edu/abs/2022arXiv221202572P} {p. arXiv:2212.02572}

\bibitem[\protect\citeauthoryear{{Reardon} et~al.,}{{Reardon}
  et~al.}{2023}]{PPTAGWReardonEtAl:2023}
{Reardon} D.~J.,  et~al., 2023, \mn@doi [\apjl] {10.3847/2041-8213/acdd02},
  \href {https://ui.adsabs.harvard.edu/abs/2023ApJ...951L...6R} {951, L6}

\bibitem[\protect\citeauthoryear{{Rhook} \& {Wyithe}}{{Rhook} \&
  {Wyithe}}{2005}]{Rhook:2005}
{Rhook} K.~J.,  {Wyithe} J. S.~B.,  2005, \mn@doi [\mnras]
  {10.1111/j.1365-2966.2005.08987.x}, \href
  {https://ui.adsabs.harvard.edu/abs/2005MNRAS.361.1145R} {361, 1145}

\bibitem[\protect\citeauthoryear{{Roebber} et~al.,}{{Roebber}
  et~al.}{2020}]{RoebberEtAl:2020}
{Roebber} E.,  et~al., 2020, \mn@doi [\apjl] {10.3847/2041-8213/ab8ac9}, \href
  {https://ui.adsabs.harvard.edu/abs/2020ApJ...894L..15R} {894, L15}

\bibitem[\protect\citeauthoryear{{Sathyaprakash} \& {Schutz}}{{Sathyaprakash}
  \& {Schutz}}{2009}]{Sathyaprakash:2009}
{Sathyaprakash} B.~S.,  {Schutz} B.~F.,  2009, \mn@doi [Living Reviews in
  Relativity] {10.12942/lrr-2009-2}, \href
  {https://ui.adsabs.harvard.edu/abs/2009LRR....12....2S} {12, 2}

\bibitem[\protect\citeauthoryear{{Sesana}}{{Sesana}}{2013}]{Sesana:2013}
{Sesana} A.,  2013, \mn@doi [\mnras] {10.1093/mnrasl/slt034}, \href
  {https://ui.adsabs.harvard.edu/abs/2013MNRAS.433L...1S} {433, L1}

\bibitem[\protect\citeauthoryear{{Sesana}}{{Sesana}}{2021}]{Sesana:2021}
{Sesana} A.,  2021, \mn@doi [Frontiers in Astronomy and Space Sciences]
  {10.3389/fspas.2021.601646}, \href
  {https://ui.adsabs.harvard.edu/abs/2021FrASS...8....7S} {8, 7}

\bibitem[\protect\citeauthoryear{{Sesana}, {Vecchio}  \& {Colacino}}{{Sesana}
  et~al.}{2008}]{SesanaEtAl:2008}
{Sesana} A.,  {Vecchio} A.,   {Colacino} C.~N.,  2008, \mn@doi [\mnras]
  {10.1111/j.1365-2966.2008.13682.x}, \href
  {https://ui.adsabs.harvard.edu/abs/2008MNRAS.390..192S} {390, 192}

\bibitem[\protect\citeauthoryear{{Sesana}, {Gair}, {Berti}  \&
  {Volonteri}}{{Sesana} et~al.}{2011}]{Sesana:2011}
{Sesana} A.,  {Gair} J.,  {Berti} E.,   {Volonteri} M.,  2011, \mn@doi [\prd]
  {10.1103/PhysRevD.83.044036}, \href
  {https://ui.adsabs.harvard.edu/abs/2011PhRvD..83d4036S} {83, 044036}

\bibitem[\protect\citeauthoryear{Skilling}{Skilling}{2006}]{SkillingNestedSampling:2006}
Skilling J.,  2006, \mn@doi [Bayesian Analysis] {10.1214/06-BA127}, 1, 833

\bibitem[\protect\citeauthoryear{{Spallicci}}{{Spallicci}}{2013}]{Spallicci2013}
{Spallicci} A. D.~A.~M.,  2013, \mn@doi [\apj] {10.1088/0004-637X/764/2/187},
  \href {https://ui.adsabs.harvard.edu/abs/2013ApJ...764..187S} {764, 187}

\bibitem[\protect\citeauthoryear{{Speagle}}{{Speagle}}{2020}]{SpeagleDynesty:2020}
{Speagle} J.~S.,  2020, \mn@doi [\mnras] {10.1093/mnras/staa278}, \href
  {https://ui.adsabs.harvard.edu/abs/2020MNRAS.493.3132S} {493, 3132}

\bibitem[\protect\citeauthoryear{{Sykes}, {Middleton}, {Melatos}, {Di Matteo},
  {DeGraf}  \& {Bhowmick}}{{Sykes} et~al.}{2022}]{SykesEtAl:2022}
{Sykes} B.,  {Middleton} H.,  {Melatos} A.,  {Di Matteo} T.,  {DeGraf} C.,
  {Bhowmick} A.,  2022, \mn@doi [\mnras] {10.1093/mnras/stac388}, \href
  {https://ui.adsabs.harvard.edu/abs/2022MNRAS.511.5241S} {511, 5241}

\bibitem[\protect\citeauthoryear{{Veitch} \& {Vecchio}}{{Veitch} \&
  {Vecchio}}{2010}]{VeitchVecchioCPNEST:2010}
{Veitch} J.,  {Vecchio} A.,  2010, \mn@doi [\prd] {10.1103/PhysRevD.81.062003},
  \href {https://ui.adsabs.harvard.edu/abs/2010PhRvD..81f2003V} {81, 062003}

\bibitem[\protect\citeauthoryear{Veitch et~al.,}{Veitch et~al.}{2022}]{cpnest}
Veitch J.,  et~al., 2022, johnveitch/cpnest: v0.11.5,
  \mn@doi{10.5281/zenodo.592884}, \url {https://doi.org/10.5281/zenodo.592884}

\bibitem[\protect\citeauthoryear{Virtanen et~al.,}{Virtanen
  et~al.}{2020}]{scipy}
Virtanen P.,  et~al., 2020, \mn@doi [Nature Methods]
  {10.1038/s41592-019-0686-2}, \href {https://rdcu.be/b08Wh} {17, 261}

\bibitem[\protect\citeauthoryear{{Volonteri}, {Habouzit}  \&
  {Colpi}}{{Volonteri} et~al.}{2021}]{Volonteri:2021}
{Volonteri} M.,  {Habouzit} M.,   {Colpi} M.,  2021, \mn@doi [Nature Reviews
  Physics] {10.1038/s42254-021-00364-9}, \href
  {https://ui.adsabs.harvard.edu/abs/2021NatRP...3..732V} {3, 732}

\bibitem[\protect\citeauthoryear{{Wang} et~al.,}{{Wang}
  et~al.}{2021}]{2021ApJ...907L...1W}
{Wang} F.,  et~al., 2021, \mn@doi [\apjl] {10.3847/2041-8213/abd8c6}, \href
  {https://ui.adsabs.harvard.edu/abs/2021ApJ...907L...1W} {907, L1}

\bibitem[\protect\citeauthoryear{Waskom et~al.,}{Waskom et~al.}{2017}]{seaborn}
Waskom M.,  et~al., 2017, mwaskom/seaborn: v0.8.1 (September 2017),
  \mn@doi{10.5281/zenodo.883859}, \url {https://doi.org/10.5281/zenodo.883859}

\bibitem[\protect\citeauthoryear{Williams}{Williams}{2021}]{nessai}
Williams M.~J.,  2021, nessai: Nested Sampling with Artificial Intelligence,
  \mn@doi{10.5281/zenodo.4550693}, \url
  {https://doi.org/10.5281/zenodo.4550693}

\bibitem[\protect\citeauthoryear{{Williams}, {Veitch}  \&
  {Messenger}}{{Williams} et~al.}{2021}]{WilliamsEtAlNessai:2021}
{Williams} M.~J.,  {Veitch} J.,   {Messenger} C.,  2021, \mn@doi [\prd]
  {10.1103/PhysRevD.103.103006}, \href
  {https://ui.adsabs.harvard.edu/abs/2021PhRvD.103j3006W} {103, 103006}

\bibitem[\protect\citeauthoryear{{Xu} et~al.,}{{Xu}
  et~al.}{2023}]{CPTAGWBXuEtAl:2023}
{Xu} H.,  et~al., 2023, \mn@doi [Research in Astronomy and Astrophysics]
  {10.1088/1674-4527/acdfa5}, \href
  {https://ui.adsabs.harvard.edu/abs/2023RAA....23g5024X} {23, 075024}

\bibitem[\protect\citeauthoryear{{Xue} et~al.,}{{Xue}
  et~al.}{2021}]{2021PhRvL.127y1303X}
{Xue} X.,  et~al., 2021, \mn@doi [\prl] {10.1103/PhysRevLett.127.251303}, \href
  {https://ui.adsabs.harvard.edu/abs/2021PhRvL.127y1303X} {127, 251303}

\makeatother
\end{thebibliography}

\appendix

\section{Posterior distributions on the population hyper-parameters}
\label{sec:iptaResults}

In this Appendix we show full corner-plots for the population hyper-parameters for the two models of the MBHB populations described in Sections~\ref{sec:models} and~\ref{sec:PTA}. 

\begin{figure*}
\centering
\includegraphics[width=\textwidth]{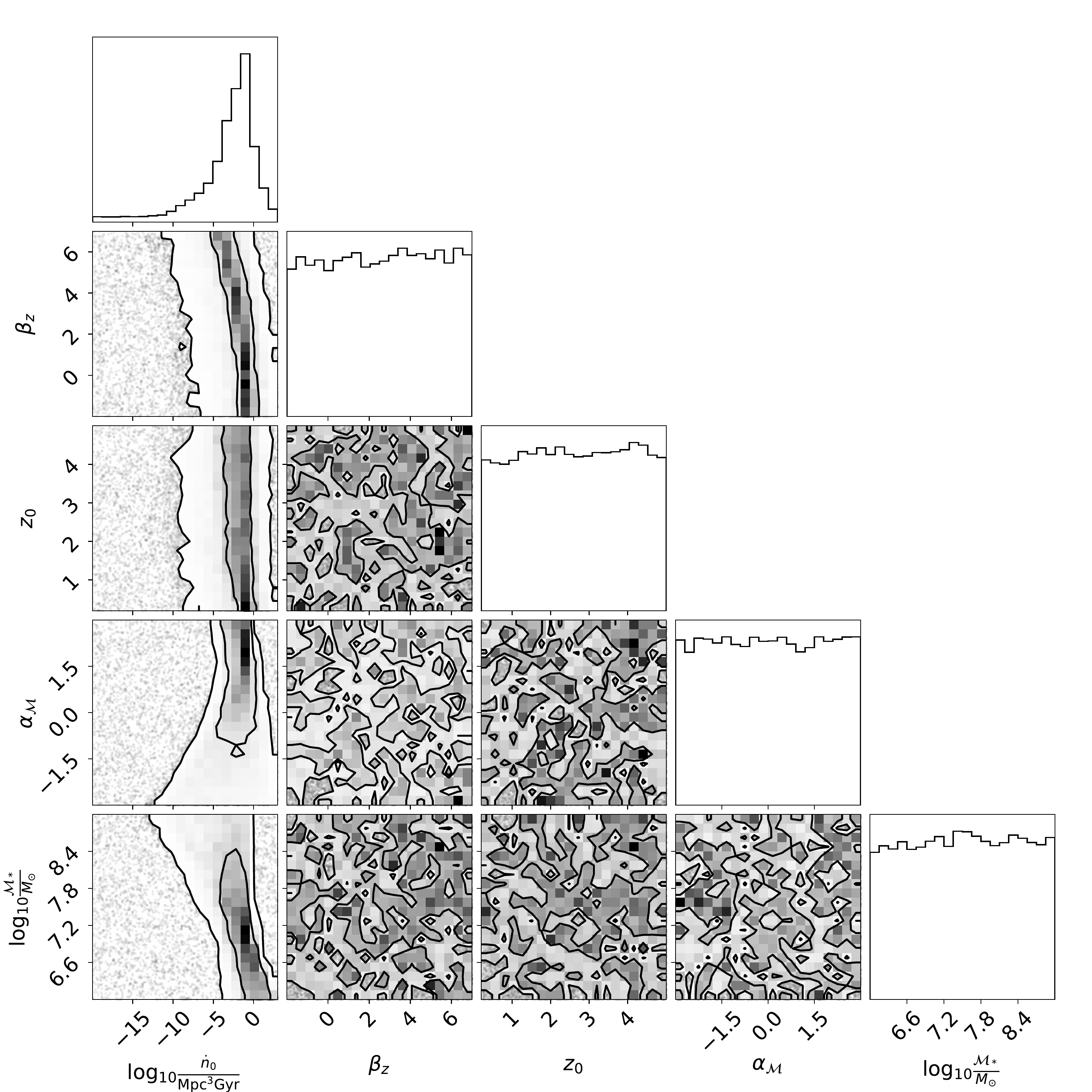}
\caption{\label{fig:agnosticModelIPTA}
Marginalised posterior distributions for the agnostic model. 
The contour plots show the two-dimensional posterior distributions for each parameter combination, where the contours indicate the central $50\%$ and $90\%$ credible regions. 
The histograms show the one-dimensional posterior distributions for each parameter.}
\end{figure*}

The marginalised posterior distributions for the five agnostic model parameters given the IPTA DR2 results are shown in Fig.~\ref{fig:agnosticModelIPTA}. 
We use flat priors in the ranges:
$\log_{10} \frac{\ndot}{{\rm Mpc}^3 {\rm Gyr}} \in [-20.0,3.0]$, 
$\betaz \in [-2.0,7.0]$, 
$z_0 \in [0.2,5.0]$, 
$\alphaM \in [-3.0,3.0]$, 
and $\log_{10} \frac{\Mstar}{\Msol} \in [10^6,10^9]$.
As in previous analysis, the only constraint from the agnostic model is on $\ndot$. 

\begin{figure*}
\centering
\includegraphics[width=\textwidth]{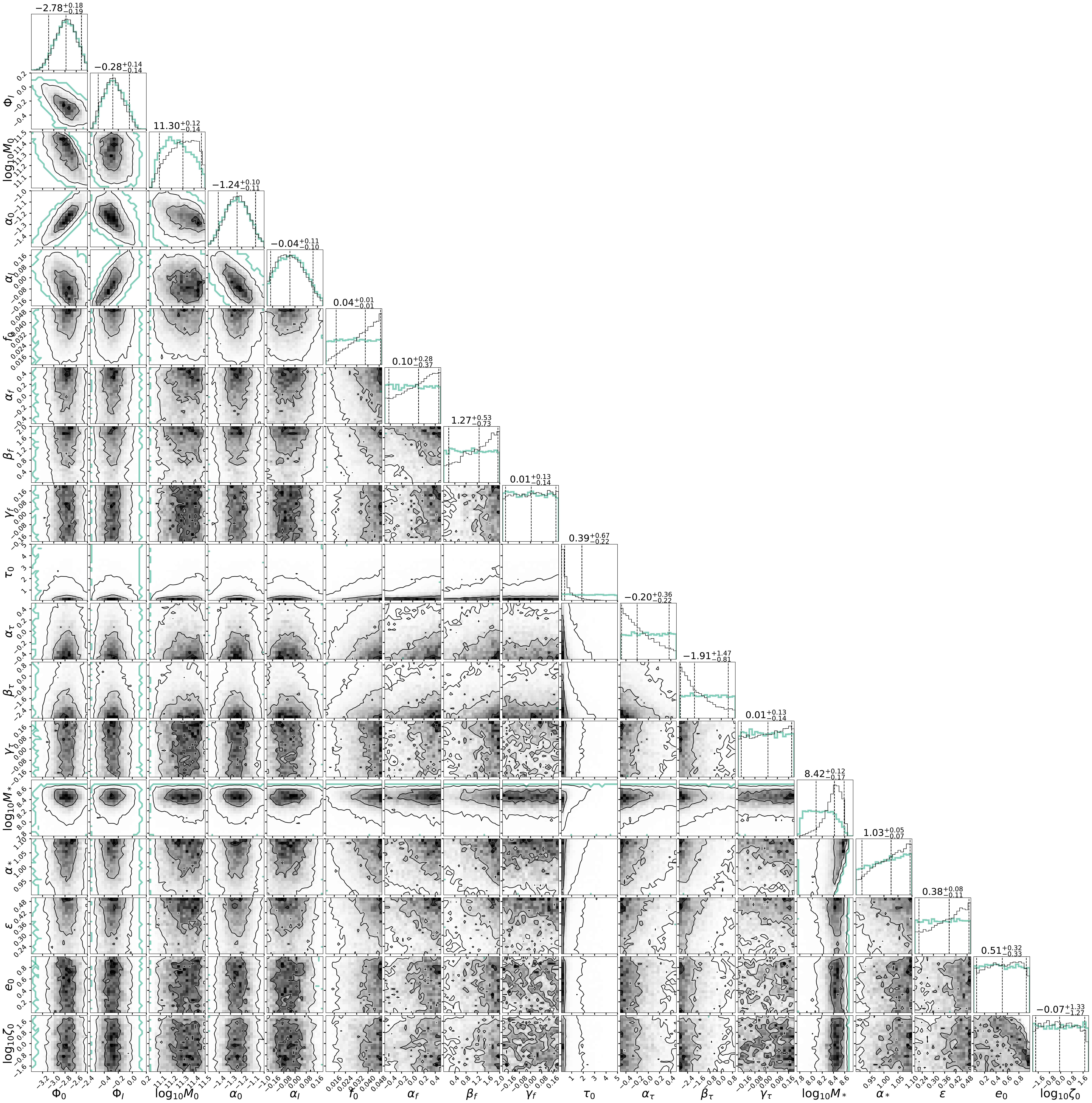}
\caption{\label{fig:astroModelIPTA}
Identical to Fig.~\ref{fig:agnosticModelIPTA} but for the astro-informed model, 
i.e., the $50\%$ and $90\%$ contours in the 2-d plots and the 5, 50, 90 percentiles in the 1-d plots. 
The green contours and histograms represent the astrophysical prior on the two-dimentional and one-dimentional distributions, respectively.}
\end{figure*}

The marginalised posterior distributions for the 18-parameter astrophysically informed model are shown in Fig.~\ref{fig:astroModelIPTA}. 
The priors are are marked in green and are identical to the extended prior ranges listed in Table I in ~\cite{ChenSesanaConselice:2019}. %

\end{document}